\documentclass{article}
\usepackage[utf8]{inputenc}
\usepackage[margin=1in]{geometry}
\usepackage{graphicx}
\usepackage[space]{grffile}
\usepackage{amsmath}
\usepackage[english]{babel}
\usepackage{mathtools}
\usepackage{amsthm}
\usepackage{amsfonts}
\usepackage{enumerate}
\usepackage{verbatim}
\usepackage{listings}
\usepackage{subcaption}
\usepackage{listings}
\usepackage{hyperref}
\usepackage{tabularx}
\usepackage{blkarray}
\usepackage{graphicx}
\usepackage{verbatim}
\usepackage{listings}
\usepackage{array}
\usepackage[refpages]{gloss}
\usepackage{amsmath}
\usepackage{amssymb}
\usepackage{courier}
\usepackage{rotating}
\usepackage{bbm}
\usepackage{color}
\usepackage{comment}
\usepackage{float,subcaption}
\usepackage{enumerate}
\usepackage{tikz}
\usetikzlibrary{calc}
\usetikzlibrary{matrix}
\usetikzlibrary{positioning}
\usepackage{epstopdf}
\usepackage{listings}	
\usepackage{color} 
\usepackage{pgfplots} 
\pgfplotsset{compat=newest} 
\pgfplotsset{plot coordinates/math parser=false} 
\usetikzlibrary{arrows.meta, positioning, quotes}
\usetikzlibrary{decorations.pathreplacing}
\usepackage{empheq}
\usepackage[]{algorithm2e}
\usepackage{braket}
\usepackage{soul}
\usepackage[numbers, square, sort&compress]{natbib}
\usepackage{ulem}
\usepackage{authblk}

\title{Using neuronal models to capture burst and glide motion and leadership in fish}
\author[1]{Linnéa Gyllingberg}
\author[2]{Alex Szorkovszky}
\author[3]{David J. T. Sumpter}
\affil[1]{{\textit{Department of Mathematics, Uppsala University, Uppsala, Sweden}}}
\affil[2]{{\textit{RITMO Centre for Interdisciplinary Studies in Rhythm, Time and Motion, University of Oslo, Oslo, Norway}}}
\affil[3]{{\textit{Department of Information Technology, Uppsala University, Uppsala, Sweden}}}
\date{}

\begin{document}
\maketitle

\begin{abstract}
While mathematical models, in particular self-propelled particle (SPP) models, capture many of the observed properties of large fish schools, they do not always capture the interactions of smaller shoals. Nor do these models tend to account for the observation that, when swimming alone or in smaller groups, many species of fish use intermittent locomotion, often referred to as burst and coast or burst and glide. Recent empirical studies have suggested that burst and glide movement is indeed pivotal to the social interactions of individual fish. In this paper, we propose a model of social burst and glide motion by combining a well-studied model of neuronal dynamics, the FitzHugh-Nagumo model, with a model of fish motion.  We begin by showing that the model can capture the motion of a single fish swimming down a channel. By then extending to a two fish model, where visual stimuli of the position of the other fish affect the internal burst or glide state of the fish, we find that our model captures a rich set of swimming dynamics found in many species of fish. These include: leader-follower behaviour; periodic changes in leadership; apparently random (i.e. chaotic) leadership change; and pendulum-like tit-for-tat turn taking. Unlike SPP models, which assume that fish move at a constant speed, the model produces realistic motion of individual fish. Moreover, unlike previous studies where a random component is used for leadership switching to occur, we show that leadership switching, both periodic and chaotic, can be the result of a deterministic interaction.  We give several empirically testable predictions on how fish interact and discuss our results in light of recently established correlations between fish locomotion and brain activity. 
\end{abstract}

%

\section{Introduction}	
A wide range of mathematical and computational models have been proposed to capture the collective motion of fish schools \citep{partridge1982structure, aoki1982simulation, reynolds1987flocks, huth1992simulation, czirok1997spontaneously, czirok1999collective, gregoire2003moving}. These models are often referred to collectively as self-propelled particle models (SPPs), with the name coming from a seminal paper by Vicsek et al.\ in 1995 \citep{vicsek1995novel}. The Vicsek model assumes that each fish (or particle) moves with constant speed, while its direction is updated at each time step to be closer to the average direction of individuals within its neighbourhood. A noise term is added to model uncertainty or error in the fish's direction. Variations of this canonical model extend it to include repulsion, attraction and other social interactions \citep{couzin2002collective, gregoire2004onset,strombom2011collective,romenskyy2017body}.

These models reproduce many of the observed properties of large fish schools \citep{tunstrom2013collective, romenskyy2017body}, but don't always capture the interactions of smaller shoals. One difference is that the assumption that the fish moves with constant speed is not typical of the swimming behaviour for many species of fish. For example, zebrafish \citep{kalueff2013towards}, koi carps \citep{wu2007kinematics}, guppies \cite{herbert2017predation}, cod \citep{videler1982energetic}, red nose tetra fish \citep{li2021burst} and many other species swim by alternating between accelerated motion and powerless gliding \citep{weihs1974energetic}. Even other animals, such as spiders, beetles and lizards have his type of intermittent locomotion \citep{kramer2001behavioral}. For swimming animals, intermittent motion is often referred to as burst and coast or burst and glide. We use the latter term in what follows. Models of burst and glide behaviour have been actively developed in the biomechanics community for decades \citep{weihs1974energetic, videler1982energetic, blake1983functional, fish1991burst, drucker1996use, akoz2017unsteady, floryan2017forces, paoletti2014intermittent}.  It has been shown that there are energetic advantages of such movements, when compared to constant swimming speed. The energetic cost of swimming is minimized during the glide phase, where the body is rigid and the fish decelerate due to water resistance \citep{weihs1974energetic}. Once a low velocity is reached, the burst phase starts again and the fish accelerates to a maximum velocity. The cycle is then completed and the fish glides again. 

Burst and glide movement is pivotal to detect and quantify social interactions between individual fish \citep{herbert2017predation}. Herbert-Read at al.\ observed that, when swimming in pairs, guppies respond to the other fish by accelerating their motion when the other fish is nearby. Pairs of fish do not update their speed continuously, but at discrete time points, making Vicsek type of models unsuitable for understanding the social dynamics of small shoals of guppies. Fish living in high predation areas have more pronounced acceleration and deceleration responses  \citep{herbert2017predation}. Moreover, Kotrschal et al.\ found that the high burst speed in response to neighbours evolves when subjected to artificial selection  \cite{kotrschal2020rapid}. 

Another aspect which is not included in the most common self-propelled particle models, but is found in fish schools of several species, is leadership.  Schaerf et al.\  showed that for pairs of freely exploring eastern mosquitofish, it is possible to categorize the fish into leaders and followers --- some fish takes the role of leaders and will on average spend more time in front than behind the other fish \cite{schaerf2021statistical}. In this case, one particular fish is the leader. However, other studies show that fish change leader-follower roles.  For example,  Nakayama et al.\ show that leadership in pairs of stickleback changes over time  \citep{nakayama2012initiative} and Milinksi show tit-for-tat like strategies in pairs of swimming sticklebacks when confronting a potential predator \cite{milinski1987tit}.

As we outline above, even though studies of interacting burst and glide behaviour are becoming more common, most studies of the burst and glide responses are of isolated fish, with a focus on understanding the energetics. Even though some self-propelled particle models assume variable speed on the part of the fish \citep{mishra2012collective}, they do not necessarily model burst and glide behaviour. Cavoli et al.\ made an interesting start in this direction by incorporating burst and glide behaviour into an SPP model \cite{calovi2018disentangling}. They have developed a model of two fish dynamics where the speed of individual fish is updated at discrete time points and the heading angle is updated due to attraction and alignment with the other fish. However, this model does not capture the observations from Herbert-Read et al.\ and Kotrschal et al.\ that there is a direct response in speed, and not just heading angle, when two fish interact with each other \cite{herbert2017predation, kotrschal2020rapid}. Thus, developing a simple model for speed response of fish interactions remains an open question.

In order to incorporate social information into burst-and-glide dynamics, it is necessary to consider how burst timings and amplitudes are governed in the nervous system. Continuous rhythmic patterns of movement in fish and other vertebrates are achieved by a neural circuit known as a central pattern generator, with afferent feedback as an important modulating component \cite{ekeberg1993combined}. Recent work indicates that the control of episodic bursting in larval zebrafish is similarly distributed, yet separable from the system that coordinates the bursts themselves \citep{wiggin2012episodic}. Although the exact mechanism for this control is still unclear, there is ample evidence for an internal burst/glide state that is subject to modulation by sensory input.

There is also an intriguing similarity between the burst and glide of fish and the burst and recover phases of neurons. The FitzHugh-Nagumo model \citep{fitzhugh1961impulses, nagumo1962active}, a well-known model of neuronal firing, is a simplification of the Hodgkin and Huxley's Nobel prize winning model for action potentials in neurons \cite{hodgkin1952quantitative}. The FitzHugh-Nagumo model has two variables, $V$ and $W$, which do not explicitly relate to specific chemical and electrical properties of neurons. Instead, $V$ models the membrane potential and $W$ is thought of as a recovery variable. The neuron dynamics in the FitzHugh-Nagumo model is given by the following equations: 
\begin{align} \label{FN-eq}
&\frac{dV}{dt}=V - \frac{V^3}{3}- W +c \\
&\tau \frac{dW}{dt}=V + a -bW.
\end{align}
The cubic-shaped dependence in the first equation captures the essential bistability of the membrane potential $V$ and, for certain parameter values, the system displays intermittent switching between firing and recovery, at a rate governed partially by the slow exponential decay of the recovery variable $W$. The constant $c$ represents a tonic (i.e.\ constant) input to the neuron.

An interesting property of the FitzHugh-Nagumo model, is that when two FitzHugh-Nagumo model neurons are coupled, the system displays chaotic dynamics \cite{shim2018chaotic}.  The evidence for chaos in neural mechanisms is extensive \cite{korn2003there,  guevara1983chaos, skarda1987brains}. Chaotic dynamics occur on both macroscopic and microscopic scales in brain dynamics \cite{wright1996dynamics} and in both humans and animals \cite{freeman1986eeg, rapp1985dynamics}.  Neuroscientific studies show that chaos is also essential for many neural mechanisms.  For example,  Ohgi et al.\ show that chaotic dynamics are fundamental in the learning and control of the dynamical interactions between brain, body and environment  in human infants \cite{ohgi2008time}.  

There are currently no existing self-propelled particle models with internal neural states explicitly defined, despite sensory information being increasingly a focus for collective motion studies \citep{strandburg2013visual,herbert2016understanding,larsch2018biological}.
Instead, models tend to be based on heuristic ``rules of interaction'' where the ability to turn sensory information into actions, such as averaging neighbour headings to obtain a turning angle, is assumed to be instantaneous and left out of the model. More complete biological models, on the other hand, are difficult to scale up to multiple agents, and the high complexity makes it difficult to connect specific mechanisms to collective outcomes.  However, there are intriguing starts in this direction made by O’Keeffe et al., where an SPP model is combined with the Kuramoto model \cite{o2017oscillators}.

Less direct analogies have previously been made between collective motion and activities in the brain, for example Marshall et al. draws parallels between decision-making in ants and decision-making in primate brains \cite{marshall2009optimal}, Passino et al.\ shows that some main characteristics of cognition in brains of vertebrates are also present in swarms of honey bees \cite{passino2008swarm}, and on a more general level, Riberio et al.\ discusses the connections between animal group dynamics and networks in brains \cite{ribeiro2020scale}. And there are existing models connecting sensory systems to neuronal dynamics. For example, Kuniyoshi et al.\ have presented a mathematical model coupling the sensory system in muscle spindles to neuronal dynamics in human infants \cite{kuniyoshi2006early}.  Furthermore, there are indications of correlations between different swimming activities and activities in fish brains \citep{dunn2016brain, naumann2016whole}, although as yet there is no proven link between burst and glide and neuronal bursting.  

In this article, we consider the possibility that even simple deterministic neural models are rich enough in dynamics to capture seemingly stochastic burst-and-glide motion patterns in fish, and that this can be used as a basis for self-propelled particle modelling. We begin with a two variable dynamical model, inspired by the FitzHugh-Nagumo model, one representing the internal state and the other representing speed, as a model for a single fish. We then introduce social interaction, and characterize the behaviour of a pair of fish as a function of key parameters. The aim of our study is primarily to investigate the mechanisms at work \citep{epstein2008model, bedau1999can, gyllingberg2023lost, smaldino2017models, braillard2015explanation}, rather than to compare the model in detail to data. Nonetheless, we show that the patterns of interactions created in the model mimic those observed in single and pairs of guppies. We discuss several testable hypotheses arising from our model.

\section{Single fish model} \label{sec:onefish}

We modify the Fitzhugh-Nagumo model so that recovery is governed by sensory feedback of the fish speed rather than internal dynamics.  Our model has two variables: the speed of the fish, $v$,  and the burst potential, $b$, which indicates whether the fish is bursting or gliding. We model motion along a line, so that $v(t)$ defines the motion entirely. The velocity itself is subject to propulsive force and drag: 
\begin{align}
&\frac{dv}{dt}=g(b)-kv\, .
\end{align}
Here $g(b)\geq 0$ is the propulsive force, which is a function of the internal state. To approximate the bursting impulse, we assume this function increases in a step-like manner from zero towards a positive constant at a critical value of $b$. We begin with linear drag with coefficient $k$, which is a suitable approximation for fish with low Reynolds number ($\lesssim 1000$) such as Trinidadian guppies or zebrafish (a quadratic drag  is often included in models of larger fish \cite{videler1982energetic,  blake1983functional, drucker1996use, akoz2017unsteady, floryan2017forces} and we will consider this in Section \ref{quadraticdrag}). The fish therefore alternates between gliding (i.e. decelerating) when $\frac{dv}{dt}=g(b)-kv <0 $, i.e., when $g(b)< kv$, and bursting (i.e. accelerating) when $\frac{dv}{dt}=g(b)-kv >0 $, i.e., when  $g(b)> kv$. The full model is now: 
\begin{align} 
&\frac{db}{dt}=b-\frac{b^3}{3}+c-v  \label{eq_onefish1} \\
&\frac{dv}{dt}=g(b)-kv,  \label{eq_onefish2}
\end{align}
where for $g(b)$ we choose the sigmoidal inverse tangent function: 
\begin{equation} \label{g-eq}
g(b)= a \left(\frac{\arctan(z_1 (b-b_0))}{\pi}+\frac{1}{2}\right). 
\end{equation}
This model again resembles the FitzHugh-Nagumo model, but with a sigmoidal rather than linear function for $g(b)$. In fact, although the functional forms are relatively simple, the shapes of these functions (cubic and sigmoidal, respectively, in the fast variable, and linear in the slow variable) are very common in fast-slow type neural models \citep{skinner1994mechanisms}.

The equation for the internal state $b$ has one main parameter, $c$, which controls the bursting rate and amplitude. This parameter represents the tonic control of frequency and amplitude of motion that is known to come from the brain stem \citep{grillner1985neurobiological,ekeberg1993combined}. 
\begin{figure}[H] \centering
\subcaptionbox*{}{\includegraphics[scale=0.5]{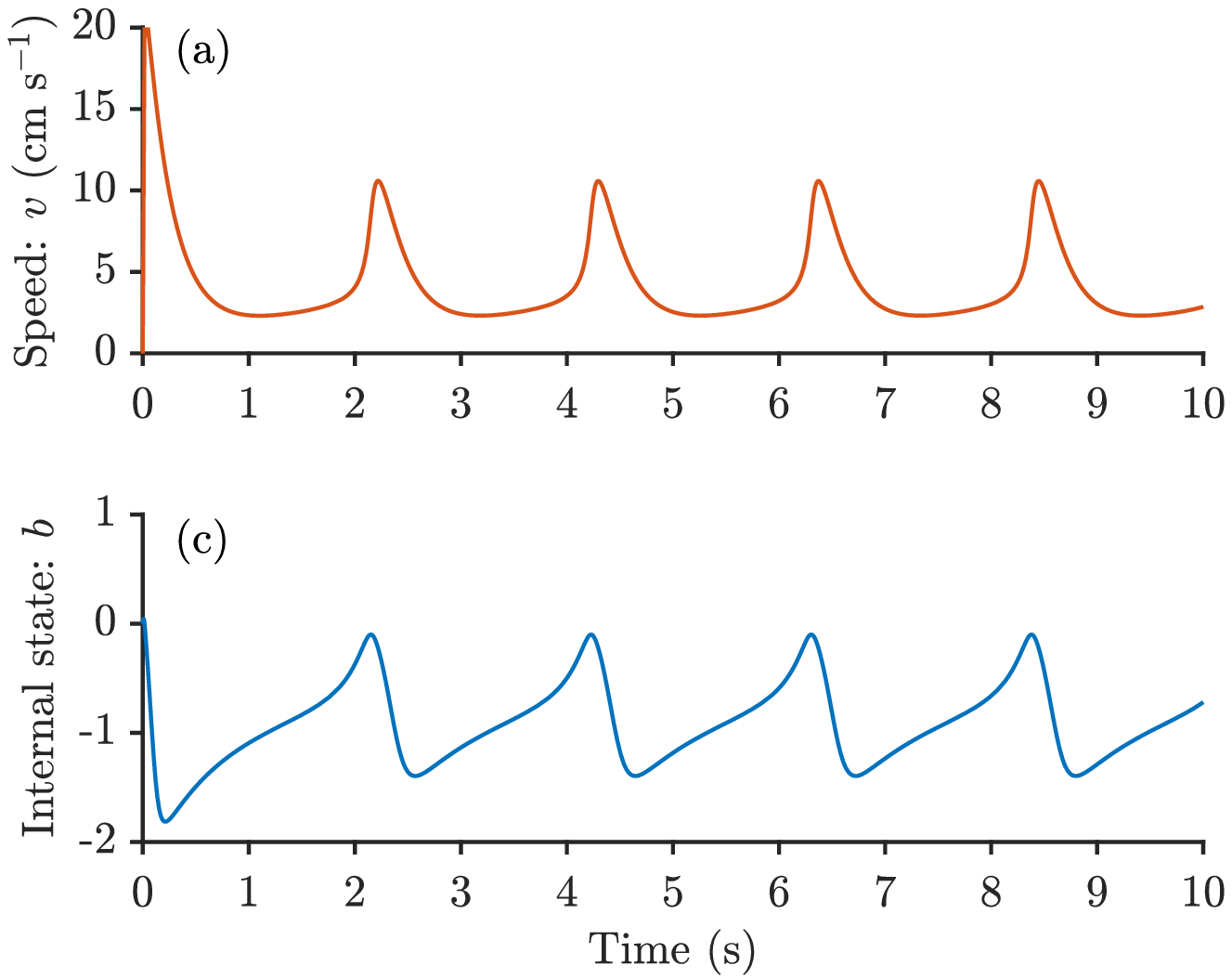}}\hspace{-0.7cm}
\subcaptionbox*{}{\includegraphics[scale=0.5]{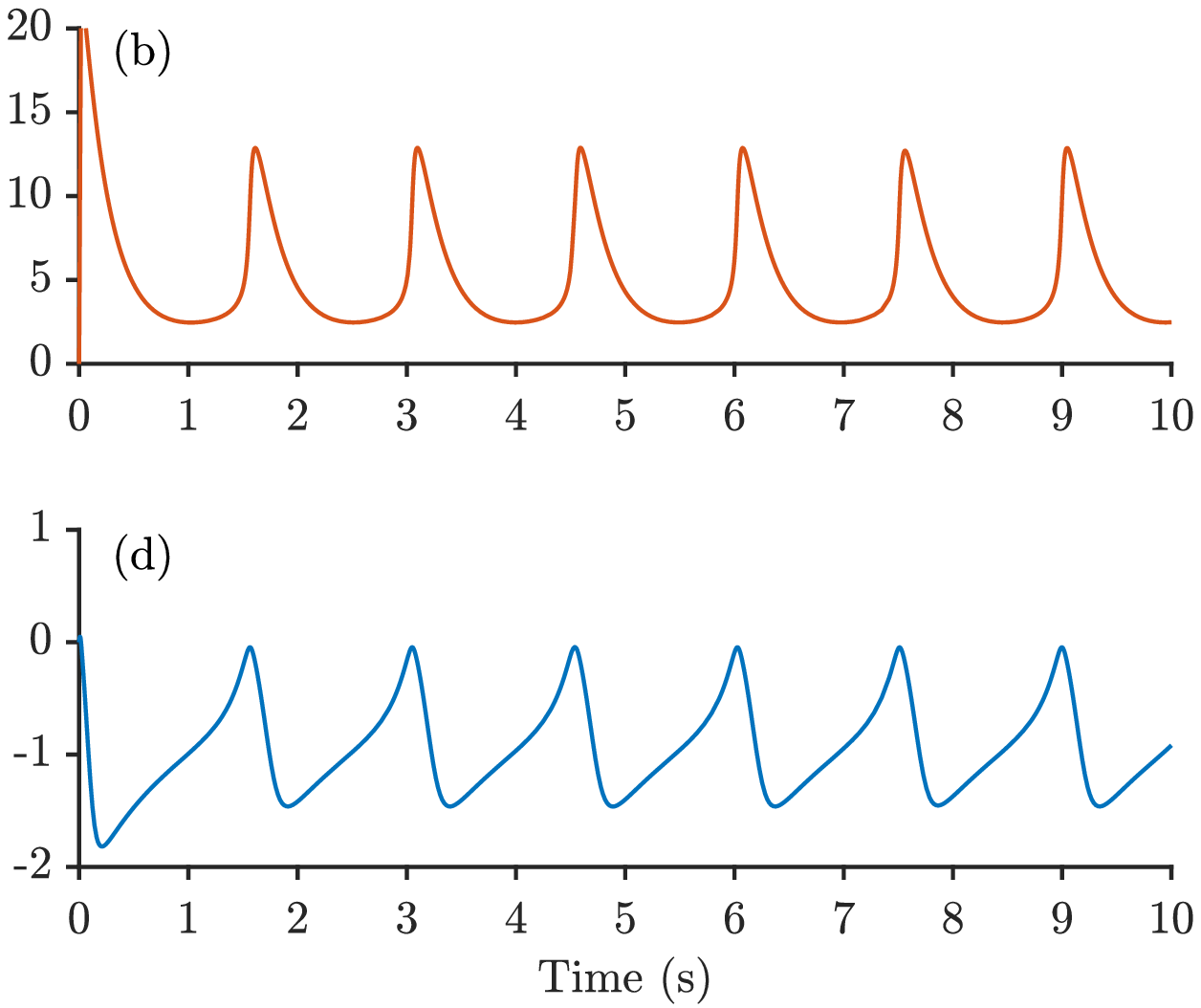}}
\caption{One fish model with $k=0.4$,  $a=15$ cm$\cdot$s$^{-2}$,  $z_1=50$ and $b_0=0$.  Panels (a) and (b) show the time series of the speed and (c) and (d) show the time series of the internal state. In (a) and (c), $c=0.95$, and in (b) and (d), $c=1$. We see that a higher value of $c$ results in a higher frequency and a higher amplitude of the speed.} \label{fig:onefish_timeseries}
\end{figure}

\begin{figure}[H] \centering
\subcaptionbox*{}{\includegraphics[scale=0.5]{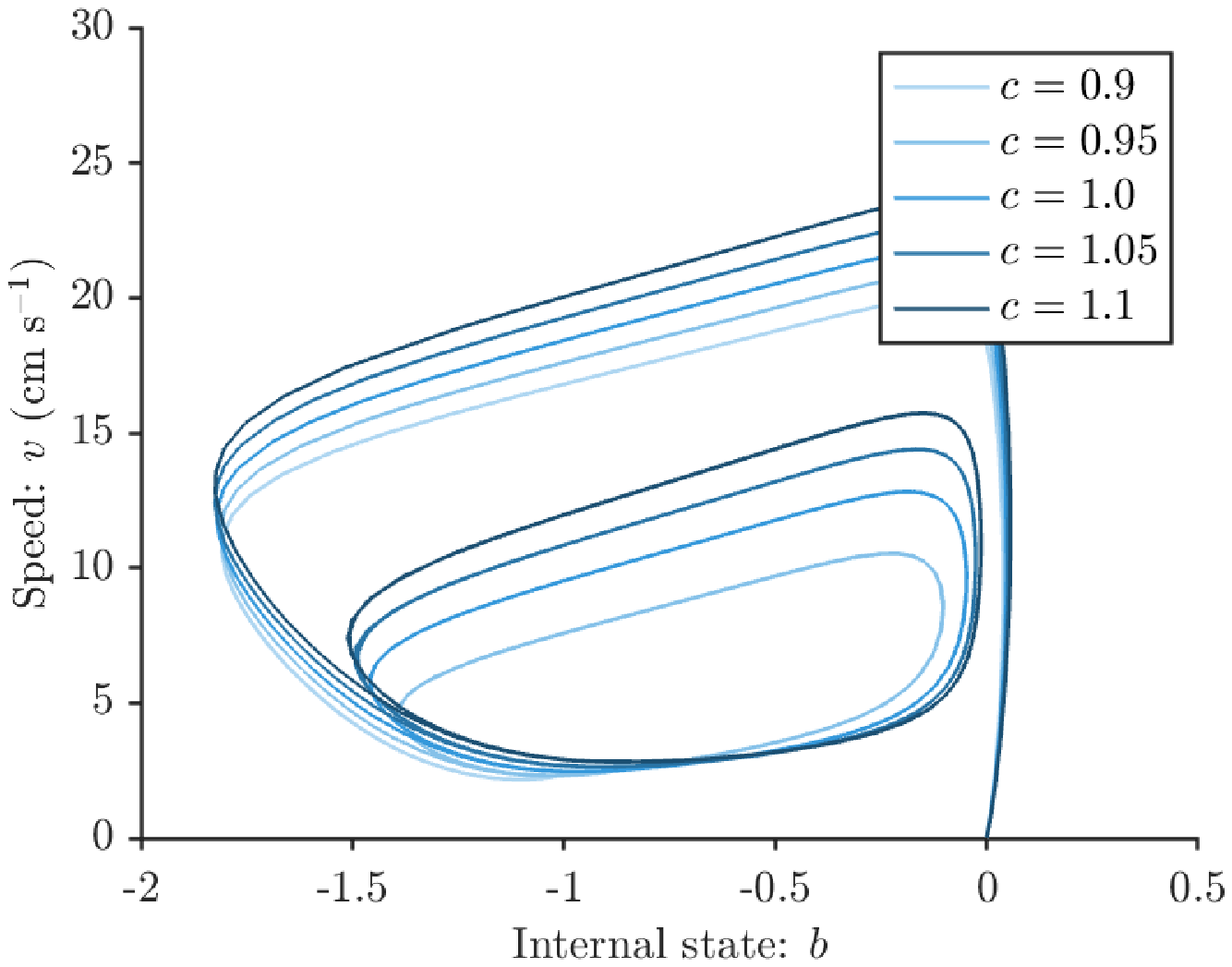}}
\subcaptionbox*{}{\includegraphics[scale=0.5]{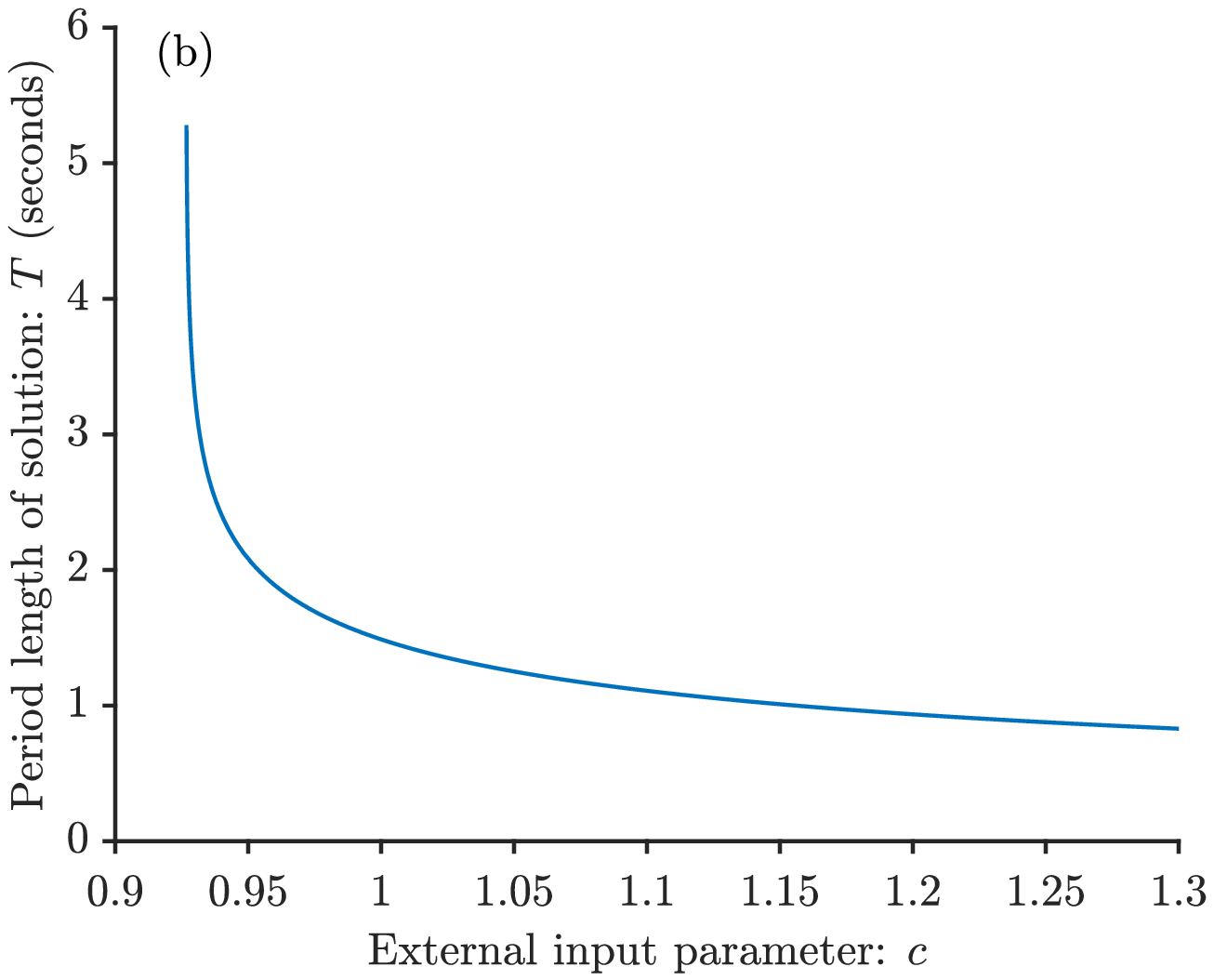}}
\caption{One fish model with $k=0.4$,  $a=15$ cm$\cdot$s$^{-2}$,  $z_1=50$ and $b_0=0$.  Figure (a) shows the phase plane for 5 different values of $c$, as shown in the legend.  For $c=0.9$ the dynamical system has a stable equilibrium,  resulting constant speed. For $c=0.95$ and larger there is a stable limit cycle. As $c$ increases the amplitude of the speed increases. Panel (b) shows the period length as a function of $c$. Increasing $c$ leads to decreasing period length and thus increasing frequency. } \label{fig:onefish}
\end{figure}

In Figure \ref{fig:onefish_timeseries}, we see the time series for the one fish model for two different values of $c$.  Both $c=0.95$ and $c=1$ result in burst and glide swimming, but $c=1$ gives a higher frequency and amplitude of the speed.  In the modelled scenario of a fish swimming down a channel, the fish accelerates about 4 cm down and then glide about 4.3 cm (for $c=0.95$).  For $c=1$ the fish accelerates about 2.3 cm, and then glides about 4.8 cm.  Figure \ref{fig:onefish} (a) shows phase plane trajectories for five different values of $c$. For $c=0.9$, there is a stable equilibrium  at  $(b^*, v^*) \approx (-1, 2.3)$,  meaning that the fish do not move with intermittent burst and glide locomotion, but with a constant speed of $\approx 2.3$ cm/s.  As $c$ is increased to $c=0.95$, there is a limit cycle, meaning there is burst and glide behaviour.  As $c$ increases further, the limit cycle increases leading to increasing amplitude of the speed.  In Figure \ref{fig:onefish} (b) we see period length of the limit cycle as a function of $c$.  We see that there is a Hopf bifurcation at $c \approx 0.927$, below which there is a stable equilibrium. On the other side of the bifurcation, the limit cycle period $T$ rapidly decreases from infinity before approaching a value on the order of one second for $c>1$. This means that the higher value of $c$, the more often the fish bursts.

\section{Two fish model} \label{sec:twofish}

\subsection{Coupling between fish}
With our single fish model  as a starting point, we now build a two fish model.  As for the single fish model, the two fish are assumed to move along a one dimensional line,  e.g., swimming up a channel, and are at the points $x_1(t)$ and $x_2(t)$ respectively at time $t$. We let $v_1(t)$ and $v_2(t)$ be the velocity of the two fish, at time $t$, and $b_1(t)$ and $b_2(t)$ be the internal state of the fish as described in Section \ref{sec:onefish}. 

We couple the equations for the two fish using a response function $f(x)$, where $x$ is the distance (negative when the other fish is behind, positive when in front) between the fish.  As we saw in the single fish model, the burst frequency and the maximum speed increases as $c$ increases.  As we want the fish to burst when it is behind another fish, the fish are coupled in the way that their internal state is affected by the distance between the fish, making $f$ act as an extra boost on $c$:

\begin{align}
&\frac{db_1}{dt}=b_1-\frac{b_1^3}{3} +c + f(x_2-x_1) - v_1\\
&\frac{db_2}{dt}=b_2-\frac{b_2^3}{3} +c + f(x_1-x_2) -v_2\\
&\frac{dv_1}{dt}=g(b_1)-kv_1 \\
&\frac{dv_2}{dt}=g(b_2)-kv_2 \\
&\frac{dx_1}{dt}=v_1 \\
&\frac{dx_2}{dt}=v_2.
\end{align}

We want the follower fish to try to catch up with the leader. Thus, for positive $x$, $f(x)$ is relatively large.  Since most fish have wide peripheral vision, the fish will also react if the other fish is just behind, meaning that $f(x)$ is also positive (but smaller) for  $x$ less than zero. It approaches zero for negative values of $x$.  To model this, we use the following sigmoid function: 

\begin{equation}
f(x_2-x_1) = d \frac{1}{1+ \exp(-z_2(x_2-x_1-x_0))}.
\end{equation}

This is a logistic function which has  $f(x)=\frac{d}{2}$ at the threshold $x_0$ and approaches $d$ as $x$ becomes large, and 0 as $x$ goes to $-\infty$. The parameter $z_2$ determines how steep the threshold response is: a high value of $z_2$ gives a steeper, more sudden, threshold response, whereas a low value gives a shallower threshold.  Since most fish have a wide peripheral vision in our model, we want a relatively less steep threshold response --  a fast switch would mean that $f(x) \approx 0$ for $x<0$, and thus entail no peripheral vision.  In our simulations, we set $x_0=15$ cm and $z_2=0.15$ cm$^{-1}$. This means the fish have responsiveness up to around 5 cm behind, but will react more strongly when the fish is swimming in front. From the single fish model we know that the burst and glide dynamics are highly sensitive to small fluctuations of the value of c. Thus $d$ takes values of size $\sim 0.01$.  Since $f(x) \to d$ as $x \to \infty$, the sigmoid function assumes that the fish would interact with the other fish infinitely far in front. This is, of course, an unrealistic assumption over large distances -- a fish would not interact with a fish 50 meters in front -- but since we are modelling short range interactions, this assumption can be used as a simplification for now.

\subsection{Temporal dynamics}

\begin{figure}[H] \centering
\subcaptionbox*{}{\includegraphics[scale=0.5]{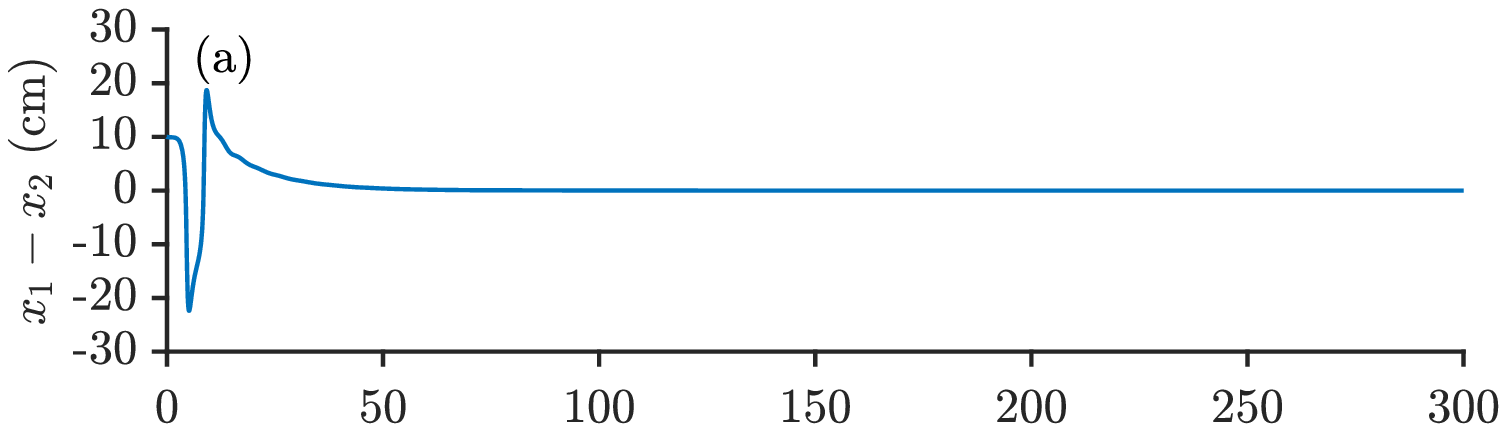}}\hspace{-0.6cm}
\subcaptionbox*{}{\includegraphics[scale=0.5]{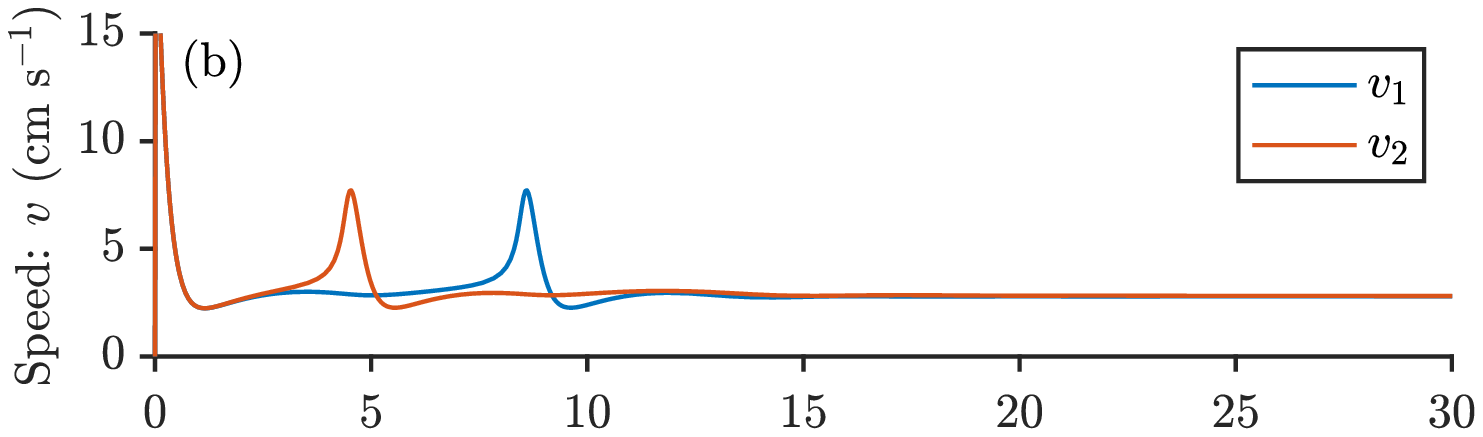}}
\subcaptionbox*{}{\includegraphics[scale=0.5]{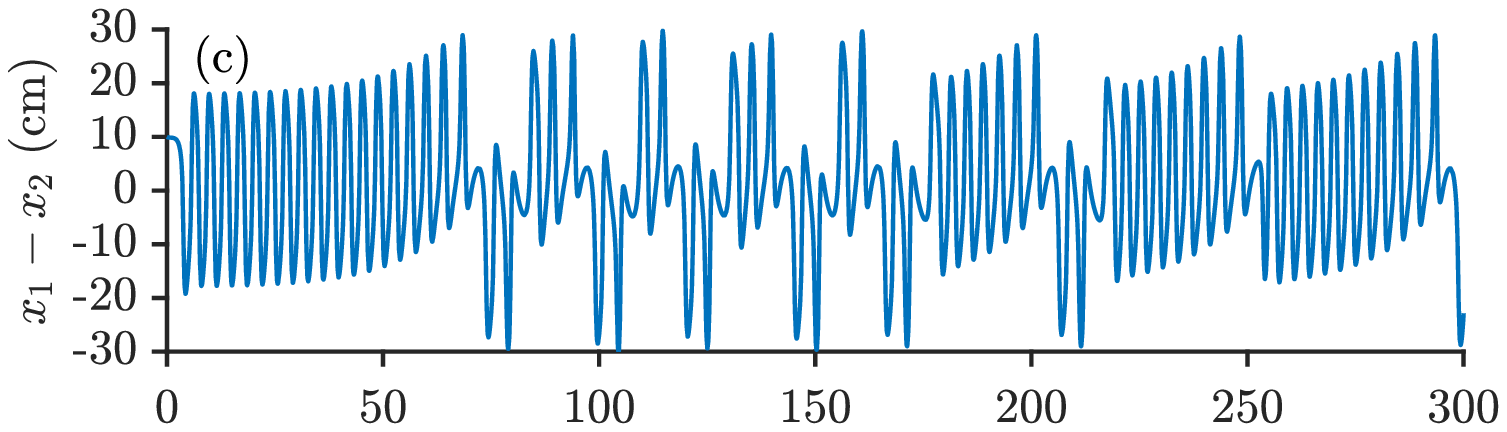}}\hspace{-0.6cm}
\subcaptionbox*{}{\includegraphics[scale=0.5]{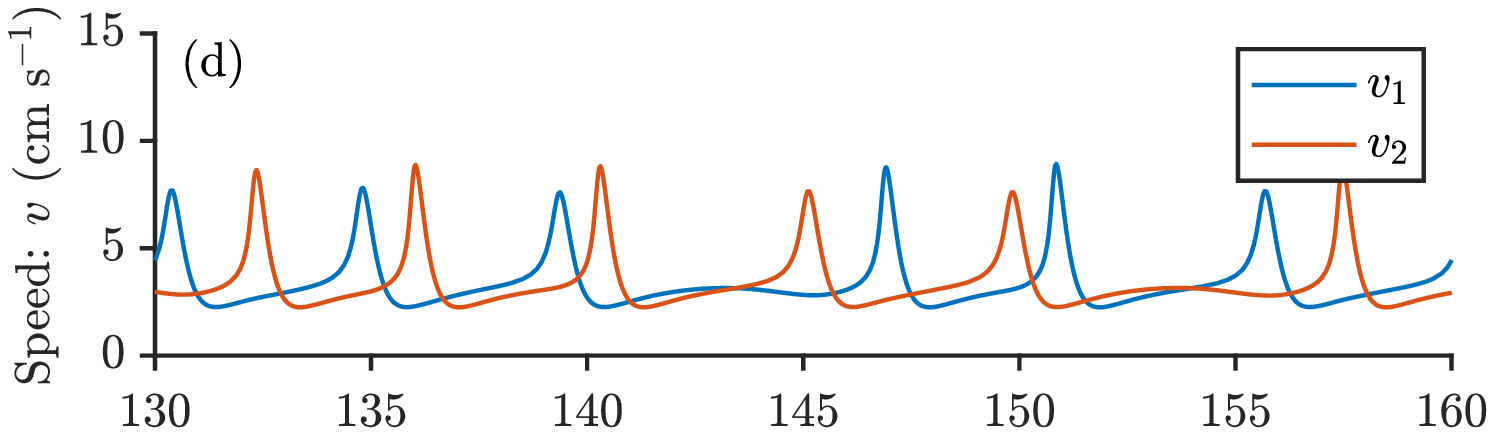}}
\subcaptionbox*{}{\includegraphics[scale=0.5]{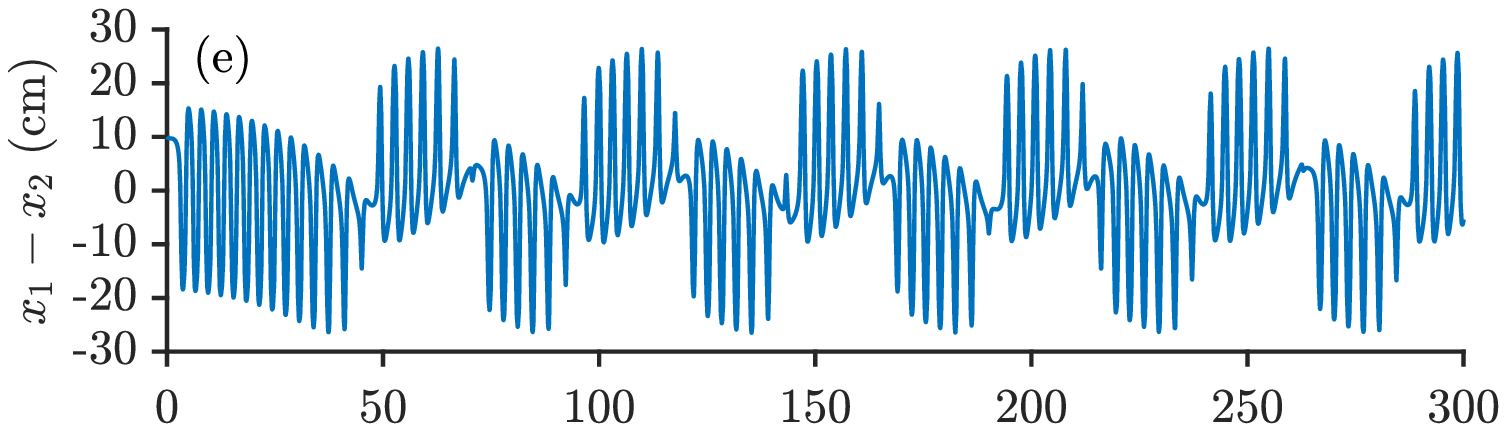}}\hspace{-0.6cm}
\subcaptionbox*{}{\includegraphics[scale=0.5]{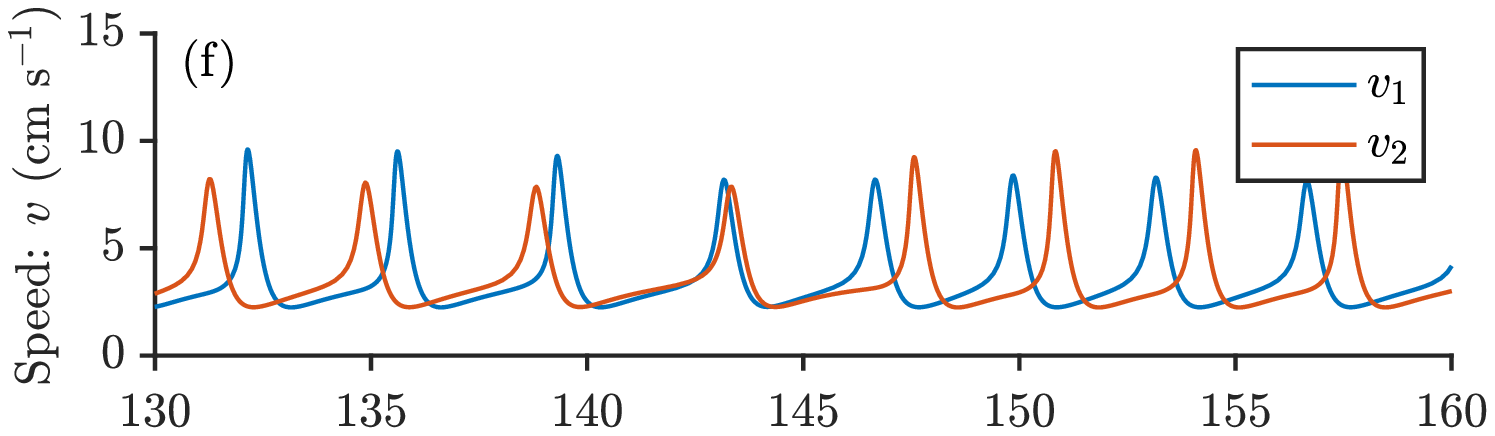}}
\subcaptionbox*{}{\includegraphics[scale=0.5]{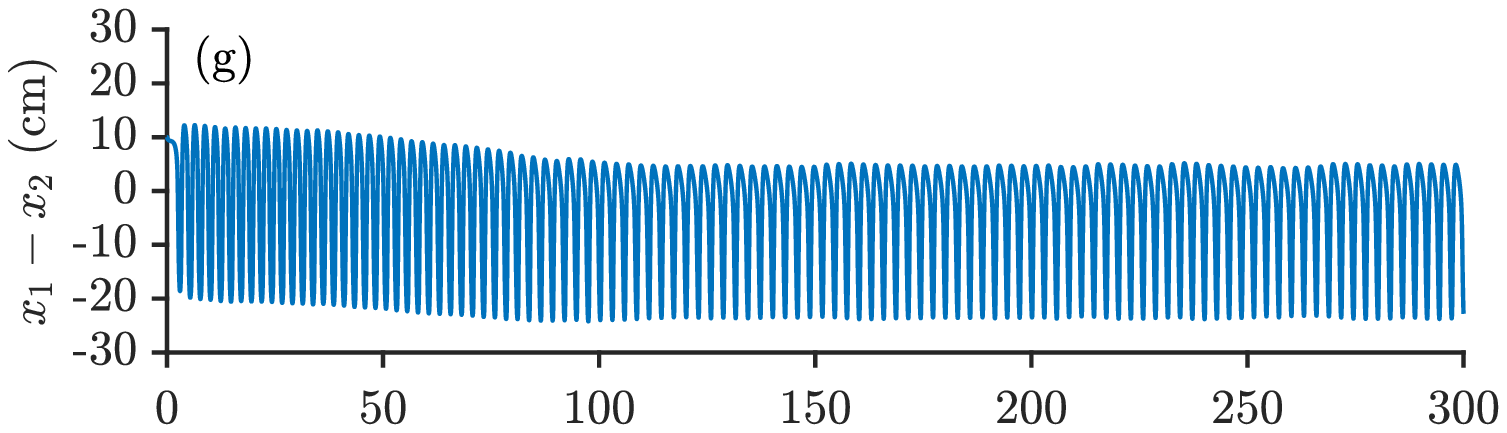}}\hspace{-0.6cm}
\subcaptionbox*{}{\includegraphics[scale=0.5]{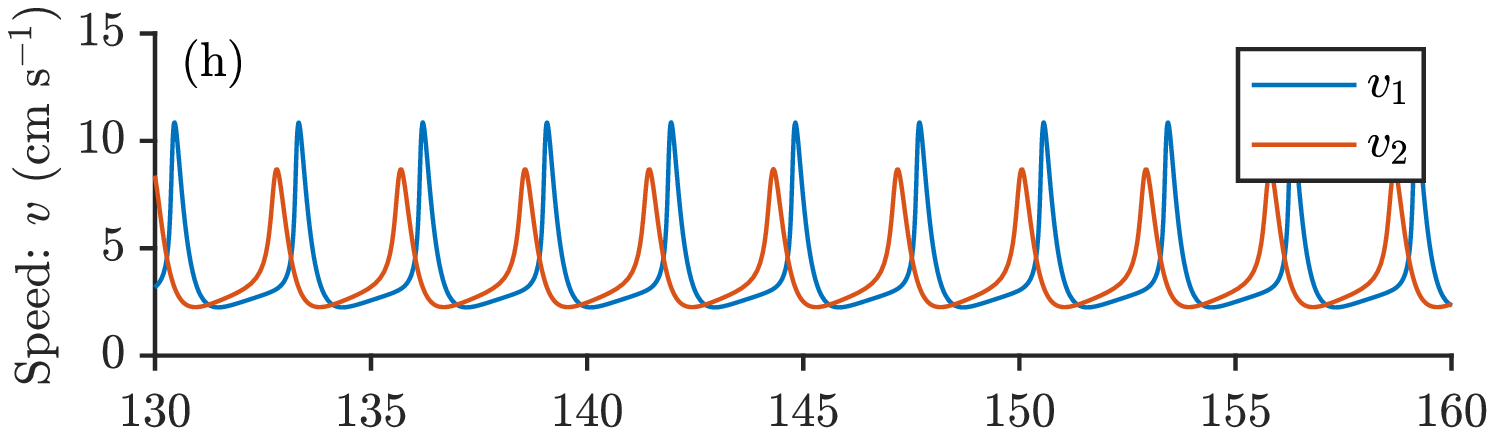}}
\subcaptionbox*{}{\includegraphics[scale=0.5]{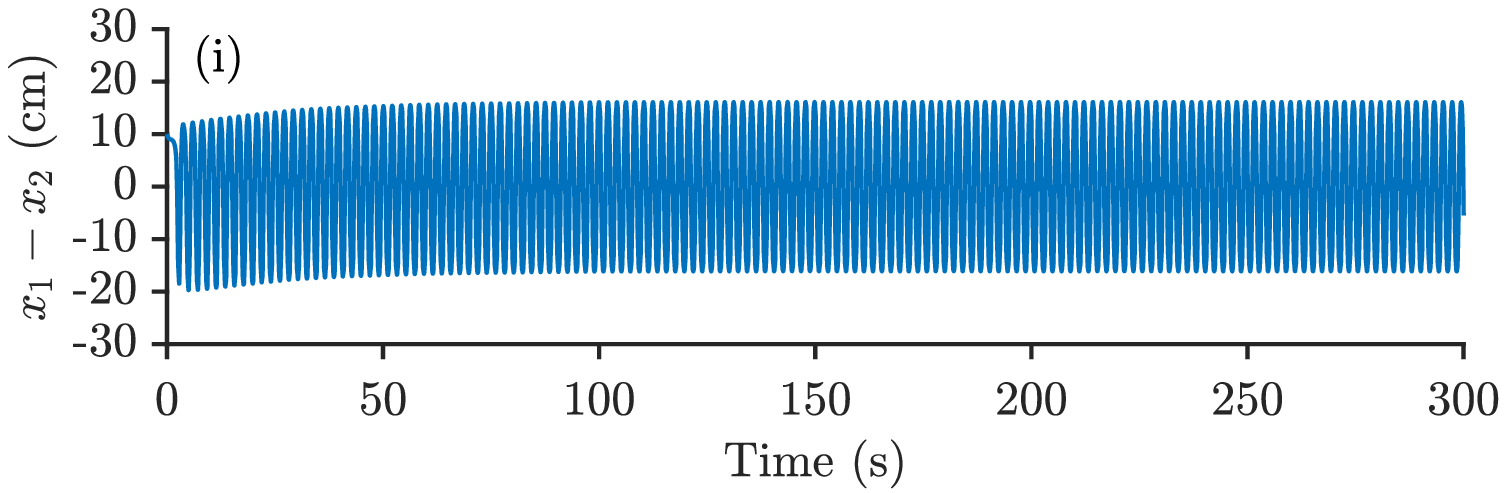}}\hspace{-0.6cm}
\subcaptionbox*{}{\includegraphics[scale=0.5]{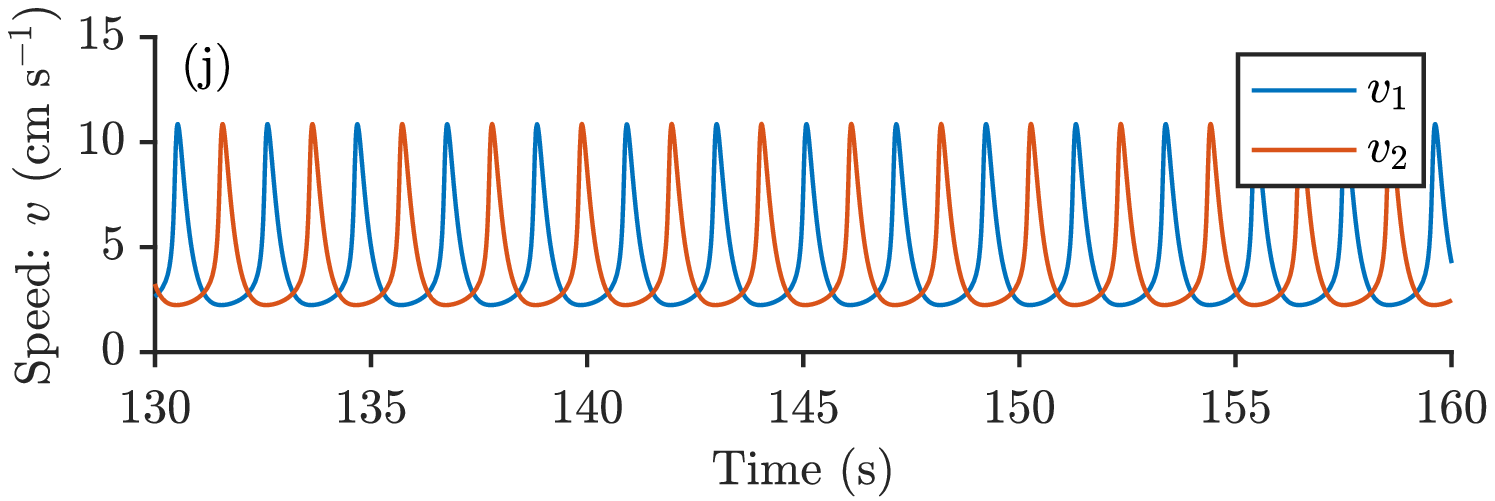}}
\caption{Two fish model with $a = 15$,  $c=0.9255$,  $z_2 =0.15$,  $x_0=15$,  $b_0= 0$, $z_1 =50$ and $k= 0.4$;.  In (a) and (b) where $d=0.005$, after one speed burst each, the fish move with constant speed and stay exactly parallel; in (c) and (d) where $d=0.01$, there is aperiodic leadership switching,  in (e) and (f) where $d=0.02$, there is periodic leadership switching, and in (g) and (h) where $d=0.05$, there is always one fish that is the leader,  and in (i) and (j) where $d=0.07$, there is no leader.} \label{fig:twofish_timeseries}
\end{figure}

We now examine the temporal dynamics of the two fish model.  The model gives rise to five different behaviours: (1) constant swimming speed with no distance between the fish, (2) aperiodic leadership switching, (3) periodic leadership switching, (4) one leader-one follower swimming, (5) no leader swimming. These are shown in Figure \ref{fig:twofish_timeseries}, where we see time series of the difference in distance ($x_1-x_2$) and the speed of the fish ($v_1$ and $v_2$) for five different values of $d$.  In Figure \ref{fig:twofish_timeseries} (a) and (b), where $d=0.005$, we see that the fish undergo one burst each, and then reach an equilibrium speed of around 2.8 cm/s. At this equilibrium the distance between the fish quickly also reaches an equilibrium at $x_1-x_2=0$. This corresponds to a steady, non-bursting swimming
on the part of the fish.

When $d$ is increased to $0.01$ (Figure \ref{fig:twofish_timeseries} (c)-(d)) both fish exhibit burst and glide motion. The fish undergo aperiodic leadership switching: the leadership of the fish alternates without any regular periodicity.  The speed of the fish (Figure \ref{fig:twofish_timeseries} (d)) is displayed from $t=130$ to $t=160$. In the beginning of this time series, fish 1 is the leader.  Fish 1 bursts about a second before fish 2, and fish 2 has a higher maximum speed in order to try to catch up. At $t \approx 143$ s there is switch in leadership and fish 2 bursts 1-2 seconds before fish 1, while fish 1 has a higher maximum speed in order to try to catch up.  After 3 bursts each, there is yet another leadership switching, and fish 1 becomes the leader again.  

Increasing $d$ even further to $0.02$ (Figure \ref{fig:twofish_timeseries} (e)-(f)), the leadership switching happens periodically: after around 7 bursts, the fish switch leader-follower roles.  There is a leadership switching at $t \approx 144$ s, where fish 1 becomes leader instead of fish 2.  In the time series of the speed of the fish (Figure \ref{fig:twofish_timeseries} (f)),  we see that before the switching, fish 1 nearly catches up with fish 2, bursting about a second after, with higher maximum speed than fish 2.  The burst by fish 1 happens closer and closer to the bursts of fish 2 on each cycle of burst and glide, and at $t\approx 143$ s, fish 1 and fish 2 burst almost simultaneously, with the same maximum speed.  After this, leadership switches so that fish 2 behind and catching up to fish 1. Now fish 2 has a higher maximum speed and burst about one second after fish 1. 

For $d=0.05$, there is always one leader: whenever the following fish starts to catch up with the leader fish, the leader burst again, which we see in Figure \ref{fig:twofish_timeseries} (g). In Figure \ref{fig:twofish_timeseries} (h), we see that the follower fish (fish 1) has a higher maximum speed than fish 2,  in order to try to catch up. Even though the follower fish has a higher maximum speed, the distance swum after one burst and glide period is approximately the same for both the follower and the leader fish.  

For $d=0.07$, there is no leader, instead the fish burst and glide with the exact same maximum speed,  but take turns in who is bursting and who is gliding (Figure \ref{fig:twofish_timeseries} (i) and (j)). This results in a distance between the fish that oscillates between $\approx-16$ cm and $\approx 16$ cm, with the mean distance between the fish equal to 0. 

\begin{figure}[H] \centering
\subcaptionbox*{}{\includegraphics[scale=0.5]{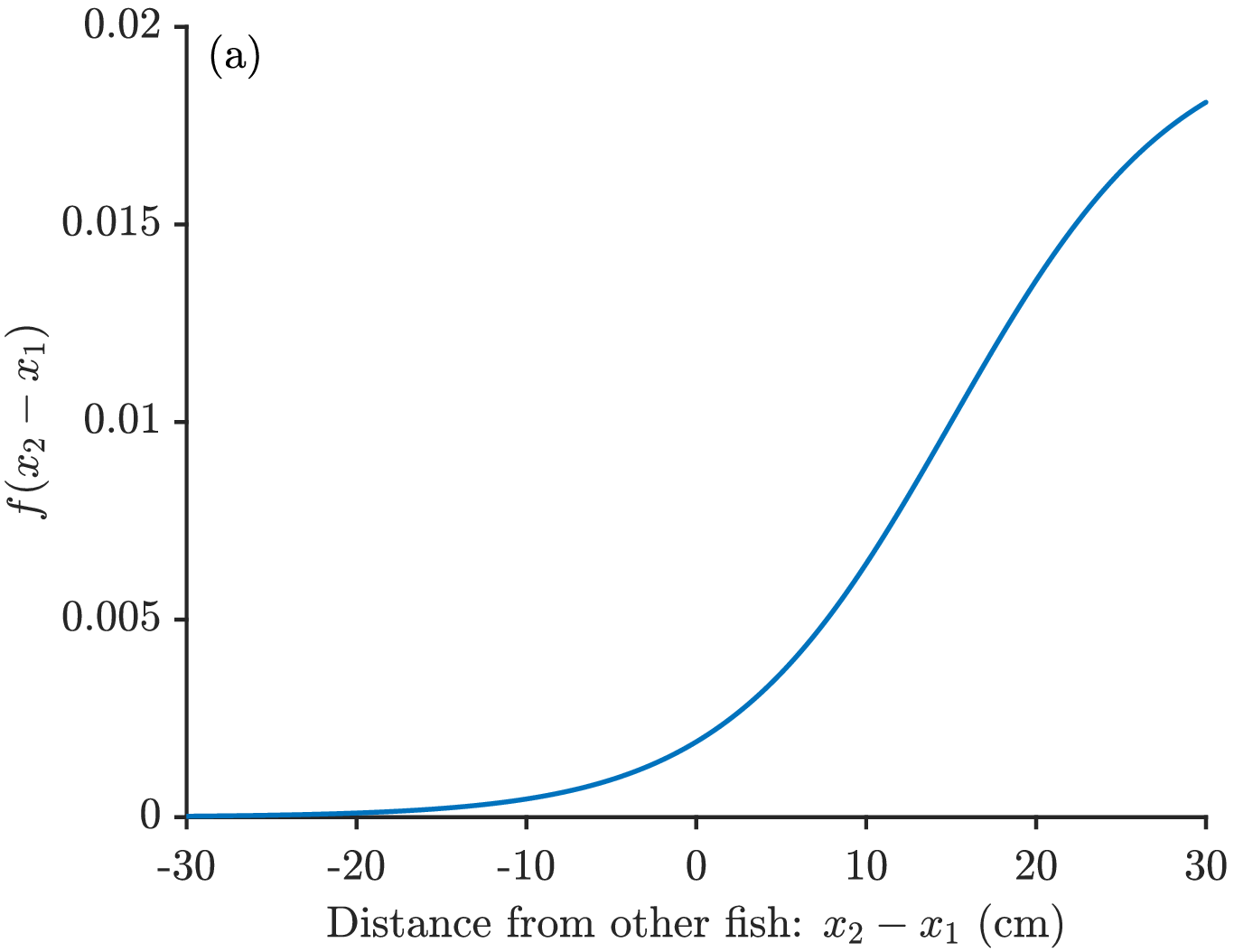}}\hspace{-0.6cm}
\subcaptionbox*{}{\includegraphics[scale=0.5]{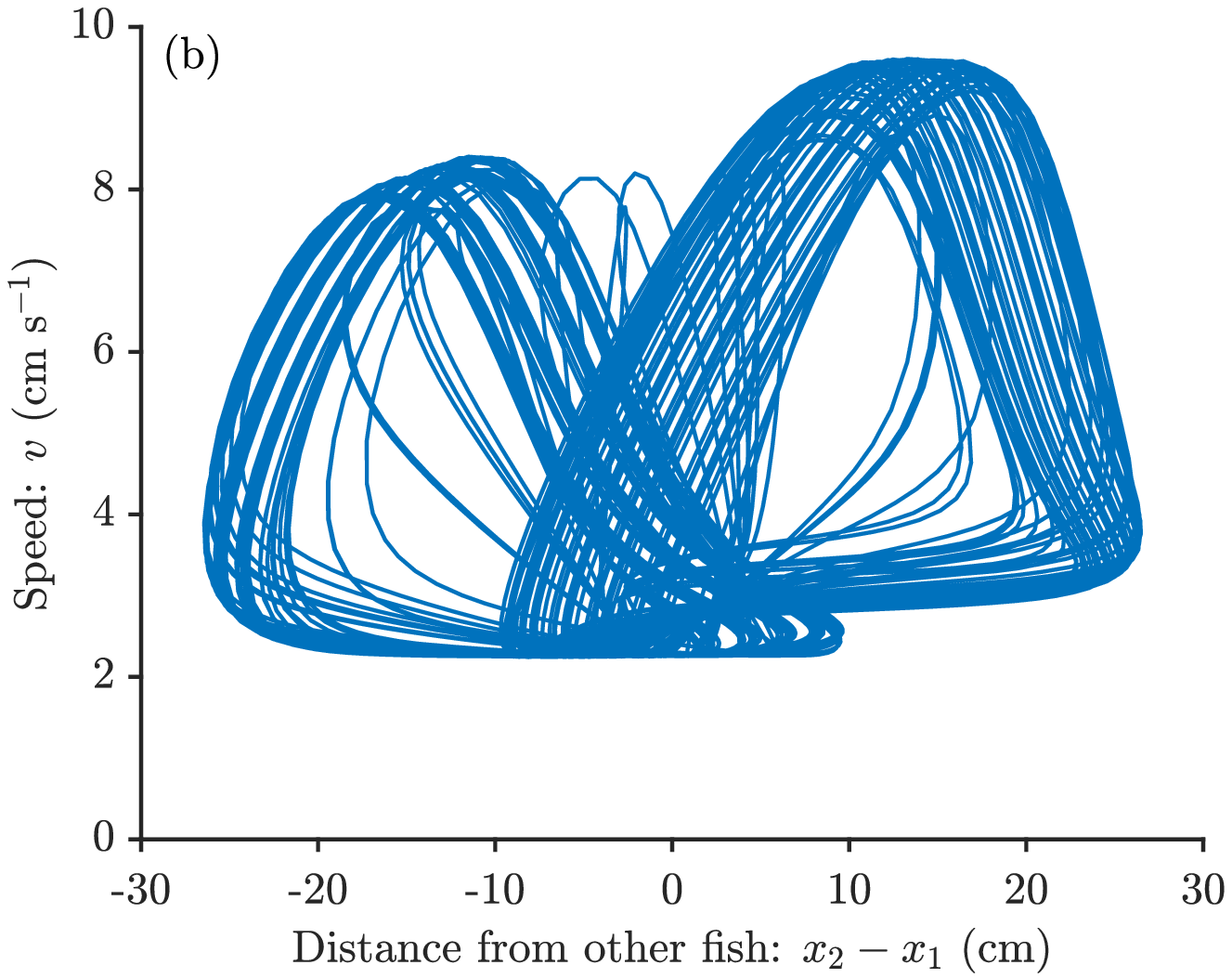}}
\caption{Two fish model with $a = 15$,  $c=0.9255$,  $z_2 =0.15$,  $x_0=15$,  $b_0= 0$, $z_1 =50$ $k= 0.4$, and $d=0.02$. In (a) the function $f(x_2-x_1)$ is shown, meaning that the x-axis shows the distance from fish 1 to fish 2 and in (b) we see the phase plane for speed of fish 1 and the distance from fish 1 to fish 2. A positive distance indicates that fish 2 is in front of fish 1, and a negative distance that fish 1 is in front of fish 2. } \label{fig:twofish_speedphase}
\end{figure}

Another way to look at these interactions is to consider the speed-distance phase plane.  In Figure \ref{fig:twofish_speedphase} (a) we see the response function $f(x_2-x_1)$ as a function of $x_2-x_1$ and in Figure \ref{fig:twofish_speedphase} (b) the speed-distance phase plane is plotted.  We notice that while the response function is asymmetrical,  a maximum speed is reached both when the fish is in front of and behind the other fish.  However, the maximum speed is higher when the fish is behind than in front, which is also what we see in Figure \ref{fig:twofish_timeseries} (f). 

\subsection{Qualitative comparison to data}

The aim of this paper is to investigate whether the model we propose qualitatively reproduces the type of motion we see between real fish. The behaviour in all of the phases, are found in experiments of pairs of fish. For example, the intermittent leadership switching in our model corresponds to stickleback swapping leadership roles \citep{nakayama2012initiative}; the constant leader-follower dynamics for higher values of the coupling parameter corresponds to the leader-follower relationship found in, for example, pairs of eastern mosquitofish \cite{schaerf2021statistical}; and the tit-for-tat like swimming is similar to the behaviour demonstrated by Milinksi \cite{milinski1987tit}.

To make a further qualitative comparison between the model and pairs of fish swimming together, we compared properties of the motion of pairs of guppies (Poecilia reticulata) to those same properties in the model. Empirical data originated from a subset of the data used in \cite{herbert2017predation} where pairs of fish were allowed to explore an open rectangular arena (1000 x 900 mm) with a depth of 45 mm for $\sim$ 16 minutes. Trials were filmed with a camera recording at 1920 x 1080 pixels at 24 frames per second. Fish in the videos were tracked using the software CTrax and tracking errors corrected with the associated Fixerrors GUI in MATLAB, providing coordinates of the centre of mass and heading (in radians) of each fish on each frame. Trajectories were analysed when fish were more than 100 mm of the corners of the arena and when the difference in heading between the pair was less than 45 degrees. Full details of the empirical methods can be found in \cite{herbert2017predation}.

%

\begin{figure}[H] \centering
\subcaptionbox*{}{\includegraphics[scale=0.47]{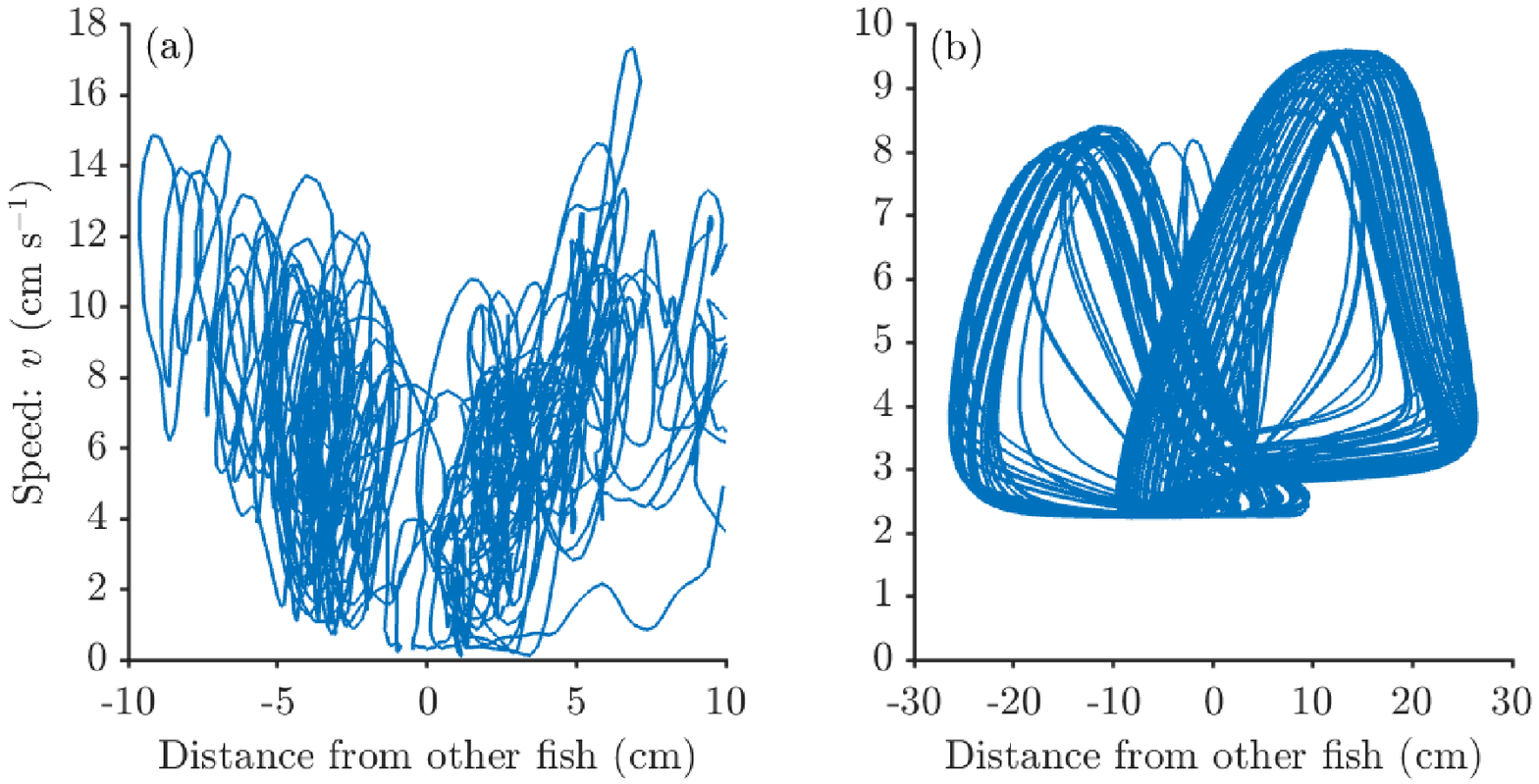}} \hspace{-0.5cm}
\subcaptionbox*{}{\includegraphics[scale=0.47]{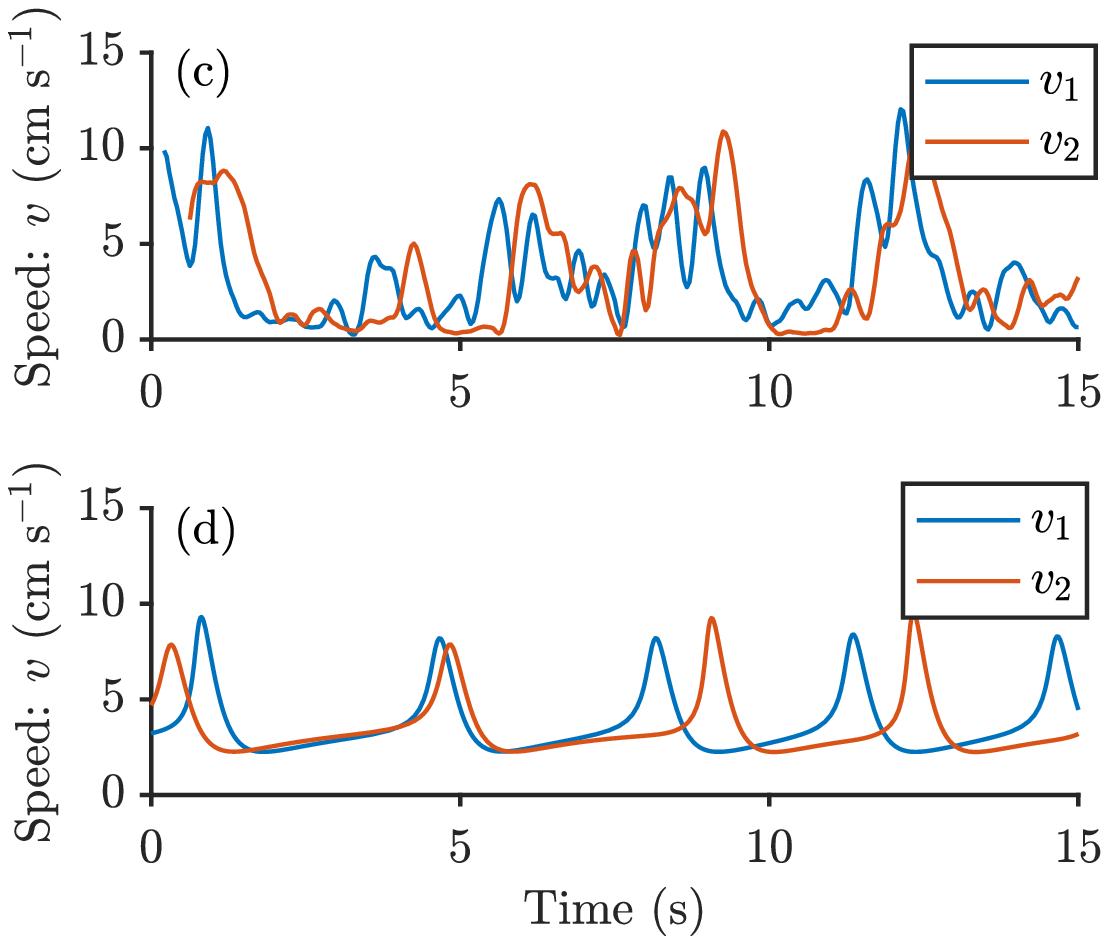}}
\caption{Model comparison with data of Trinidadian guppies ({\em Poecilia reticulata}) from \cite{herbert2017predation}.  (a) shows the phase plane of the distance between the two fish and the speed of one fish, compared to (b) the same phase plane for the model simulation for parameters set a $a = 15$,  $c=0.9255$,  $z_2 =0.15$,  $x_0=15$,  $b_0= 0$, $z_1 =50$ $k= 0.4$, and $d=0.02$.  In (c) (data) and (d) (model) we see a comparison of the speed of the two fish as a function of time for a period of 15 seconds.  Data is from the the median trial out of all female pairs in terms of activity (quantified by median speed over the trial). All time points are included where both fish are more than 100 mm away from corners of the arena, and with a difference in heading of less than 45 degrees.} \label{fig:empirical}
\end{figure}

  In Figure \ref{fig:empirical} we see a phase plane plot based on data from \cite{herbert2017predation}, compared to the model, together with a short time series from both the model and data from the same data set.  We see that the model produces a symmetrical phase plane just as the data from the experiments.  However, the minimum speed for the model is higher and the maximum speed is lower than the speed in experiments.  In the short time series comparison, we see that the data from the experiments are noisier, but have approximately the same amount of burst and glides within 15 seconds.

\subsection{Role of parameters in dynamics}
We now further investigate the dynamics for the different types of swimming behaviour. Since the first swimming behaviour,  constant swimming speed with no distance between the fish,  consists of an equilibrium point for both speed of the fish and distance between the fish, we will not investigate this swimming behaviour in our further analyses.  For the other dynamics, we will use a similar technique as first described by Lorenz \cite{lorenz1963deterministic}.  When studying what later became known as the Lorenz equations, he plotted the maximum value of his $z$ variable on consecutive journeys round the continuous time attractor. In doing so, he found that the maximum obeyed something similar to a tent map \cite{lorenz1963deterministic}, which is known to produce chaotic dynamics. 

\begin{figure}[H] \centering
\subcaptionbox*{}{\includegraphics[scale=0.5]{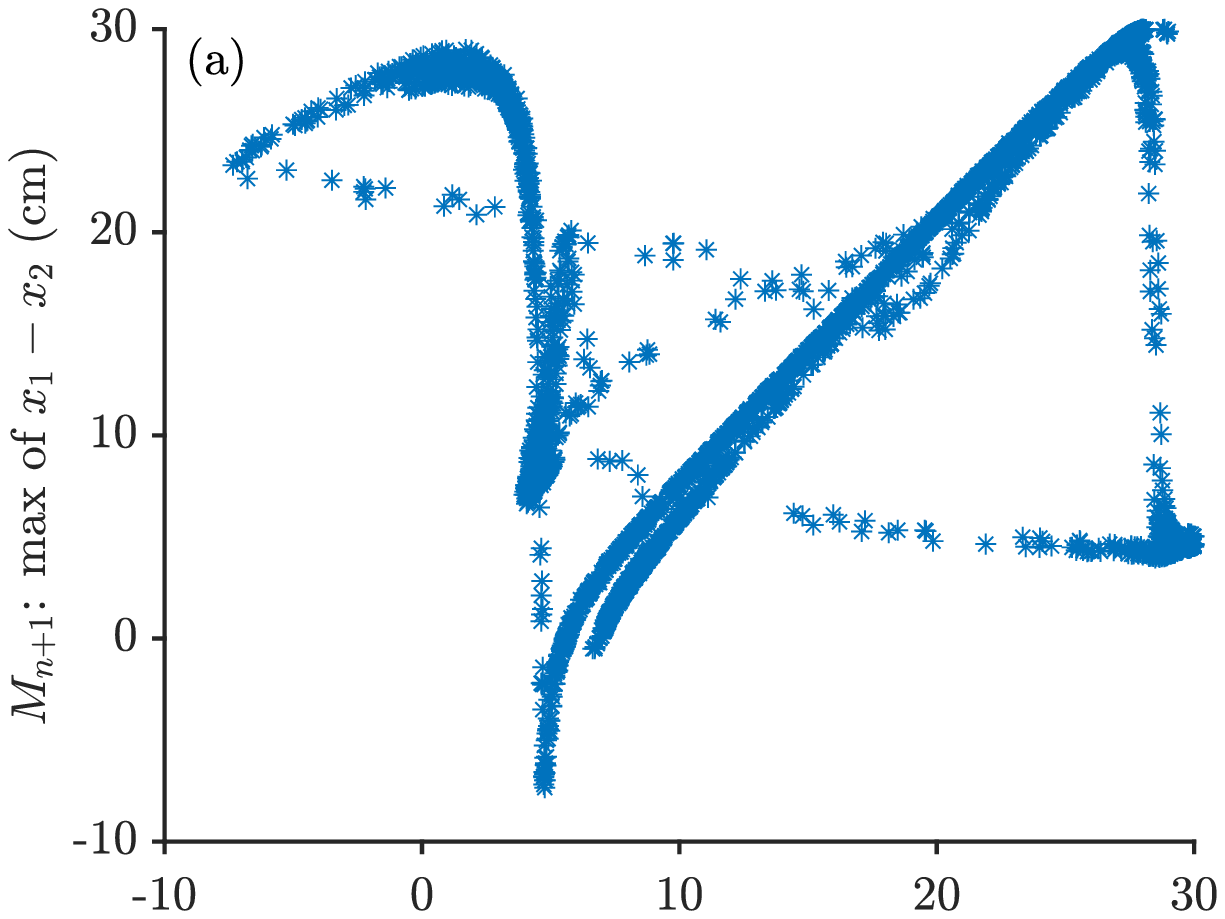}}
\subcaptionbox*{}{\includegraphics[scale=0.5]{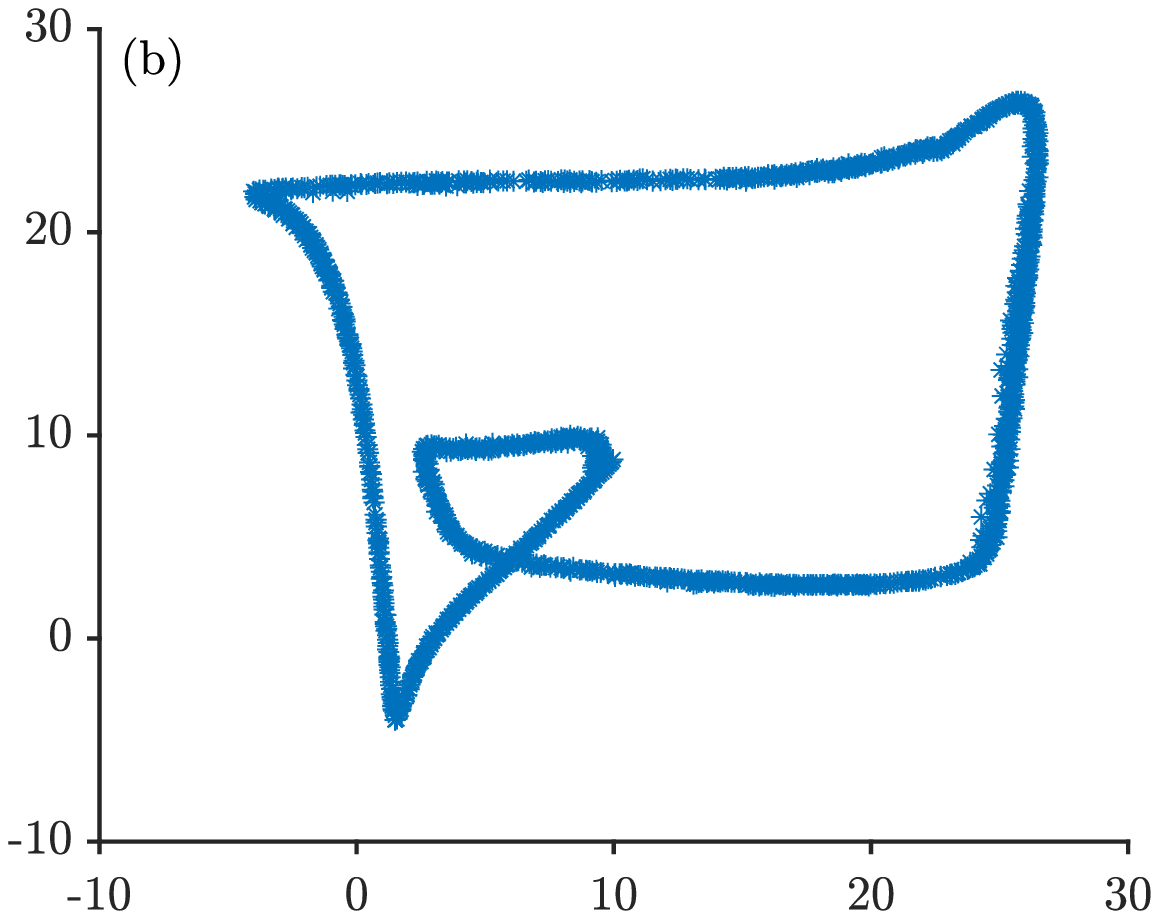}}
\subcaptionbox*{}{\includegraphics[scale=0.5]{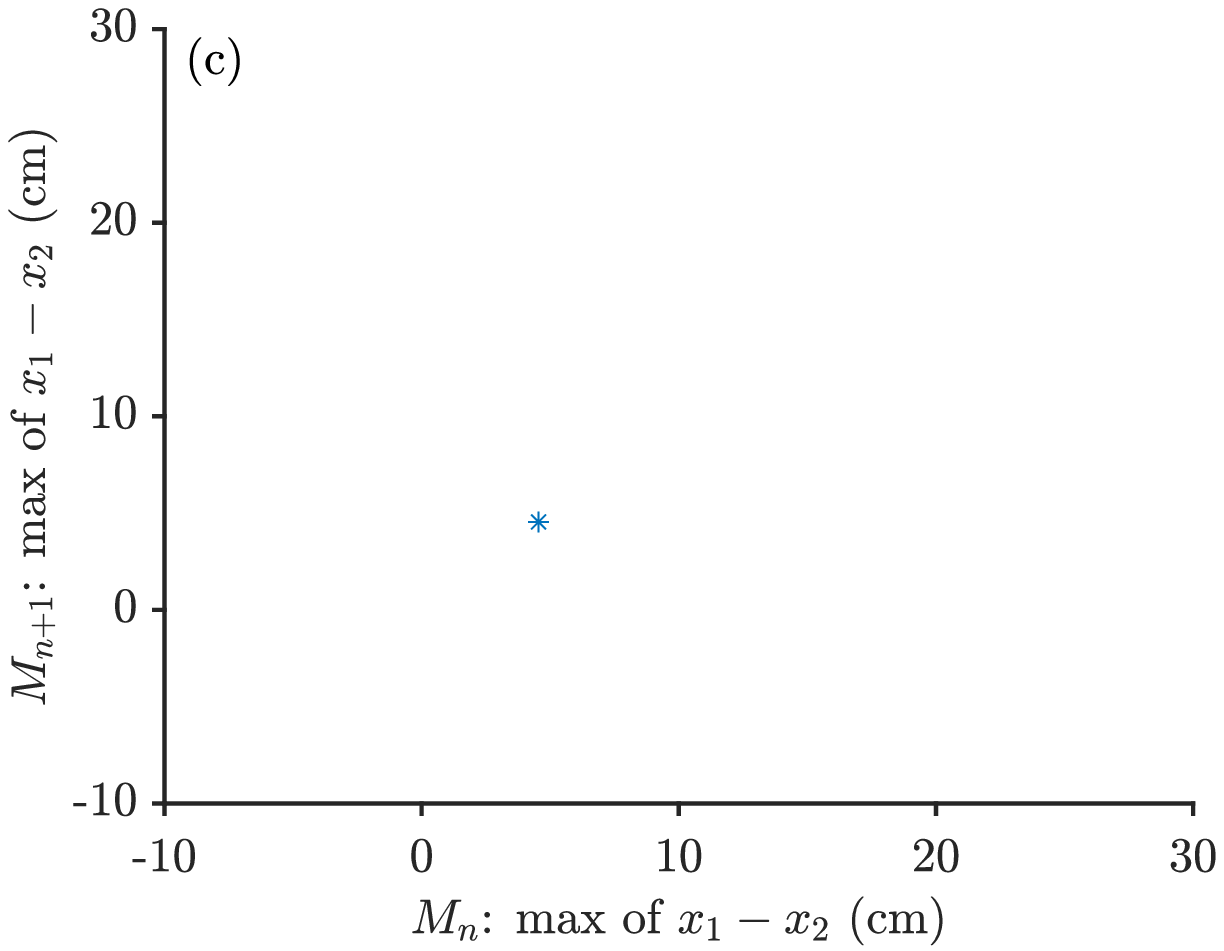}}
\subcaptionbox*{}{\includegraphics[scale=0.5]{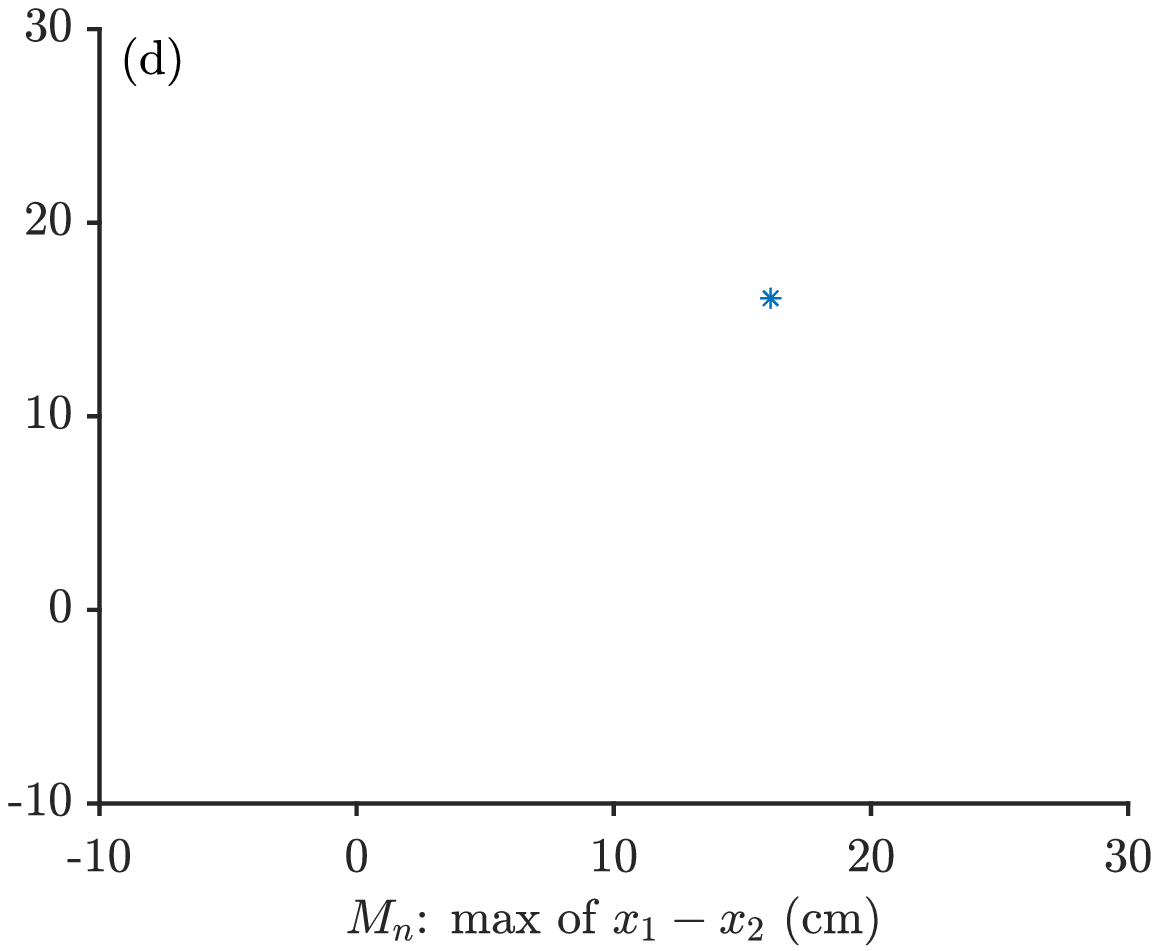}}
\caption{Lorenz map for two fish model with $a = 15$,  $c=0.9255$,  $z_2 =0.15$,  $x_0=15$,  $b_0= 0$, $z_1 =50$ and $k= 0.4$, for maximum value.  In (a) $d=0.01$,  in (b) $d=0.02$, in (c) $d=0.05$ and (d) $d=0.07$.} \label{fig:twofish_lorenz}
\end{figure}

Inspired by this approach, we find the peaks (maximum) of the difference in distance between the fish, $x$, and then plot the value of the $n$th maximum, $M_n$, against the next maximum $M_{n+1}$.  These plots are shown in Figure \ref{fig:twofish_lorenz}.  For our system, this approach does not give a tent map for any of the parameter values, but it does give a clearer impression of the dynamics. In Figure \ref{fig:twofish_lorenz} (a) we see that the Lorenz map for $d=0.01$ has a complicated structure, where as in Figure \ref{fig:twofish_lorenz} (b) when $d=0.02$, the Lorenz map shows a periodic pattern. In Figure \ref{fig:twofish_lorenz} (c) and (d), i.e. when $d=0.05$ and $d=0.07$ respectively, the maximum value have reached an equilibrium point at 4.53 and 16.1 cm, respectively.

\begin{figure}[H] \centering
\subcaptionbox*{}{\includegraphics[scale=0.5]{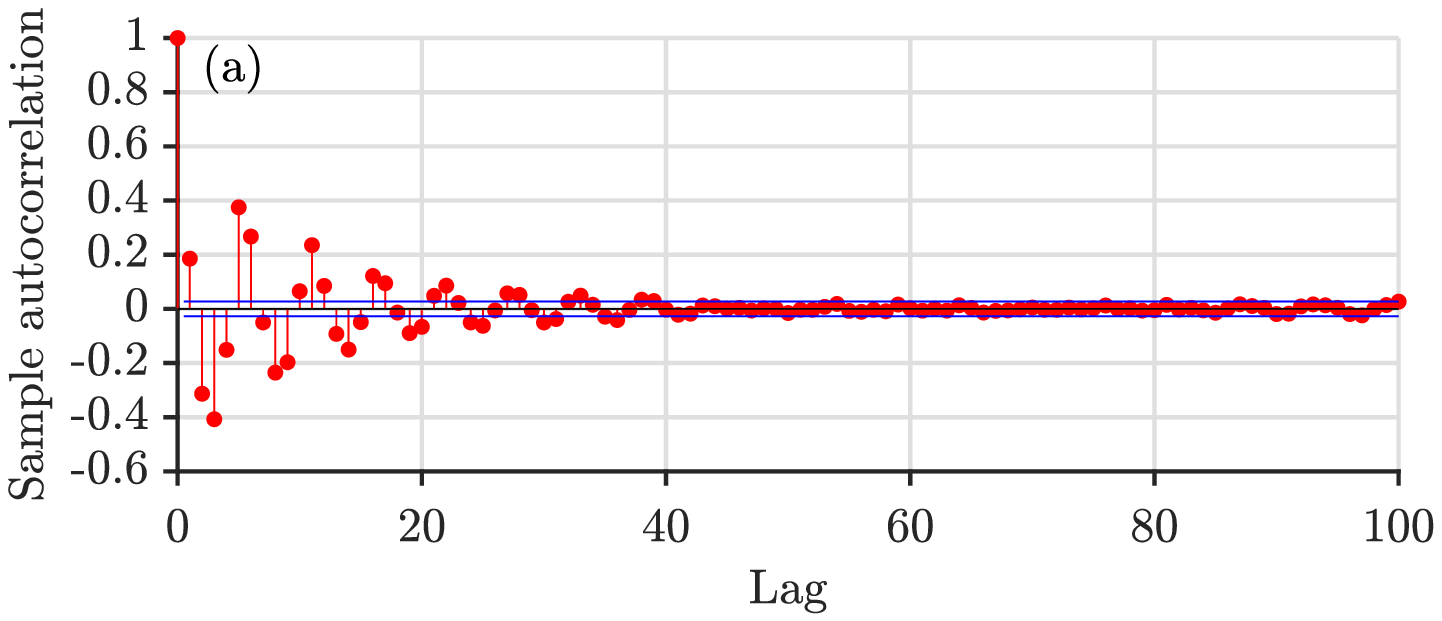}}
\subcaptionbox*{}{\includegraphics[scale=0.5]{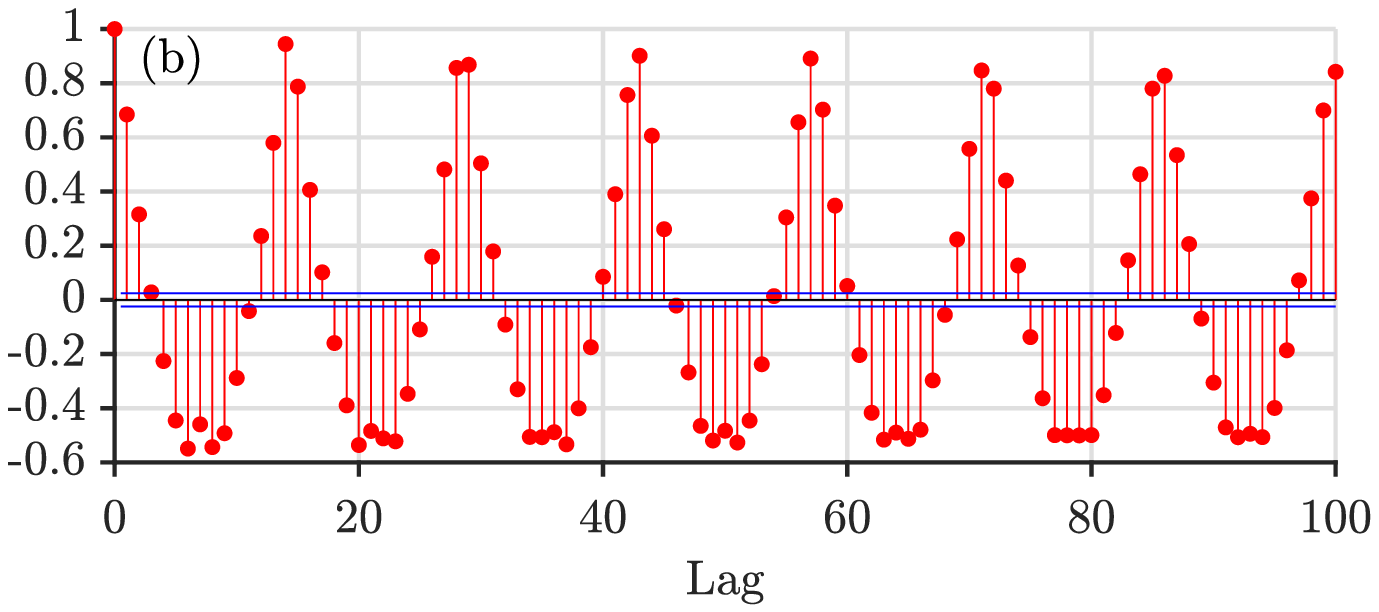}}
\caption{Autocorrelation for peaks of distance difference $a = 15$,  $c=0.9255$,  $z_2 =0.15$,  $x_0=15$,  $b_0= 0$, $z_1 =50$ and $k= 0.4$;.  In (a) $d=0.01$, and in (b) $d=0.02$.} \label{fig:autocorr_peaks}
\end{figure}

Another approach for investigating the differences between the various behaviours is to look at the autocorrelation function, ACF, for the peaks of the distance difference, $M_n$.  In Figure \ref{fig:autocorr_peaks} we plot the autocorrelation function of the peaks of distance,  $M_n$, for $d=0.01$ and $d=0.02$.  When $d=0.01$ (Figure \ref{fig:autocorr_peaks} (a)),  there is a weak periodic pattern of length 2-3 in the ACF. This is most likely due to the fact that many leadership switchings happens after 2-3 bursts, but apart from that, there is no clear periodic structure.  In Figure \ref{fig:autocorr_peaks} (b), where $d=0.02$, we see a clearly periodic pattern in the ACF.  The periodicity is of length $\approx 14$, which corresponds to leadership switching after $\approx 7$ bursts.  Since the peaks of the distance difference for both $d=0.05$ and $d=0.07$ consists of one equilibrium point, we do not analyse the ACF for these values of $d$. 

\begin{figure}[H] \centering
\subcaptionbox*{}{\includegraphics[scale=0.38]{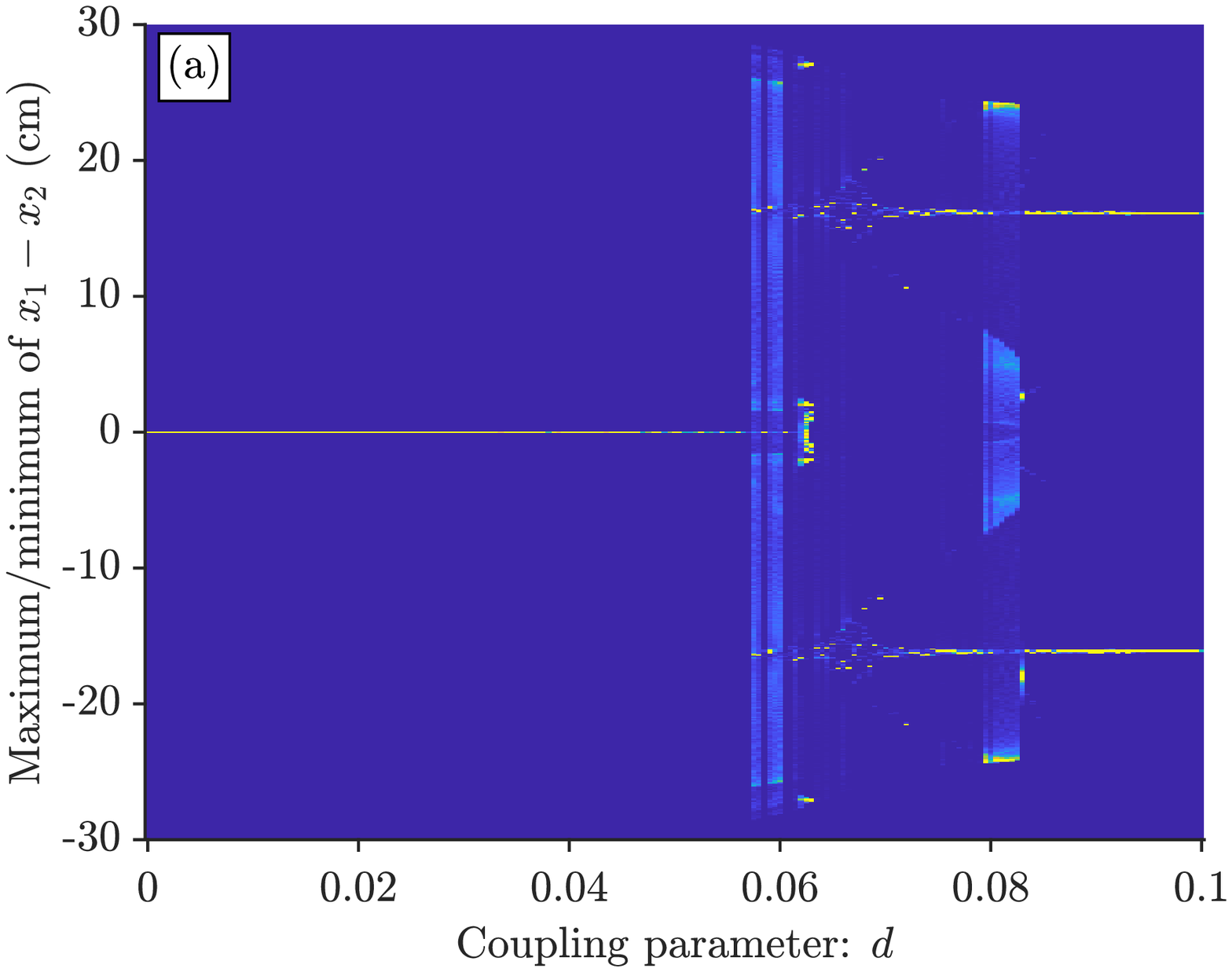}}
\subcaptionbox*{}{\includegraphics[scale=0.38]{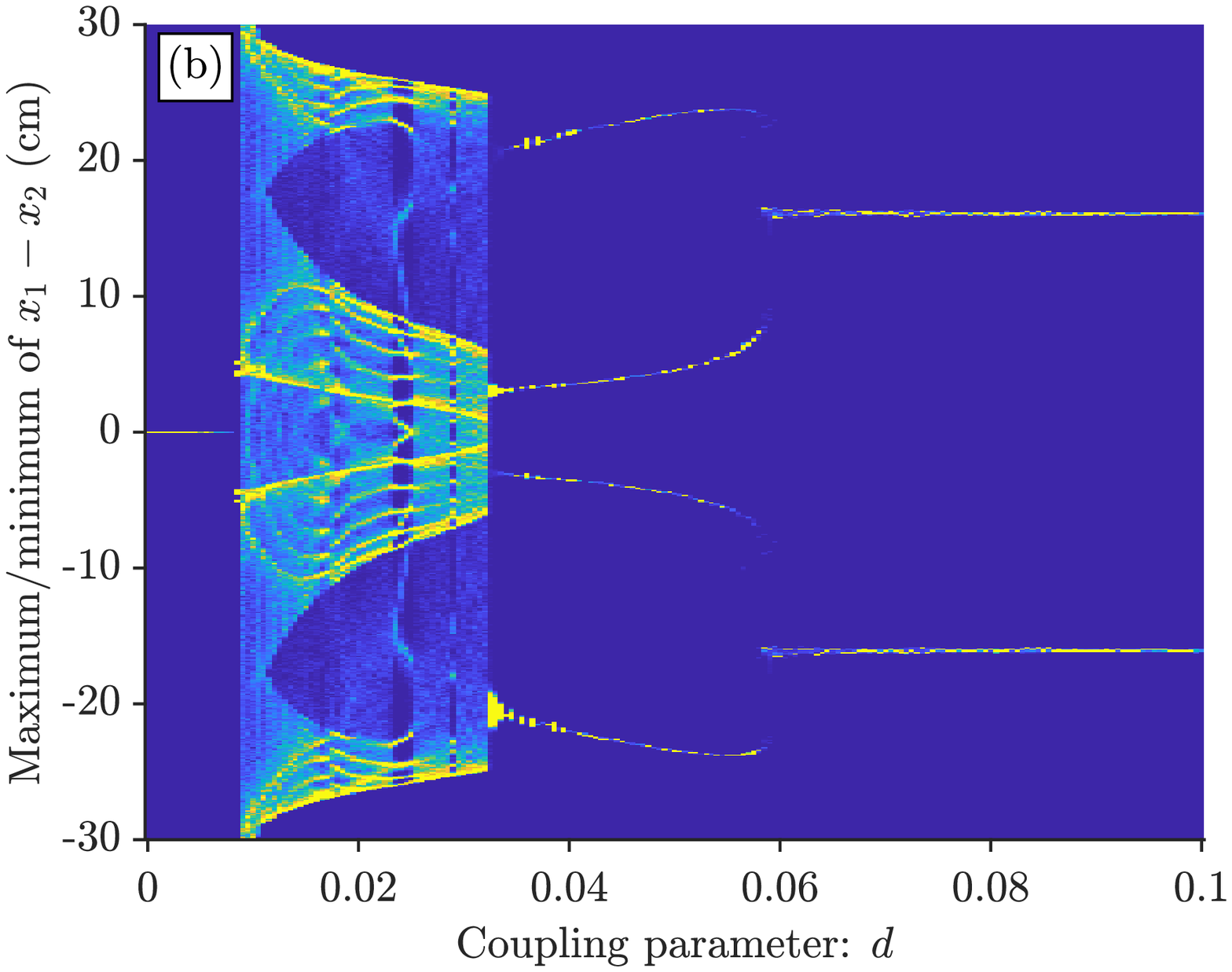}}
\subcaptionbox*{}{\includegraphics[scale=0.38]{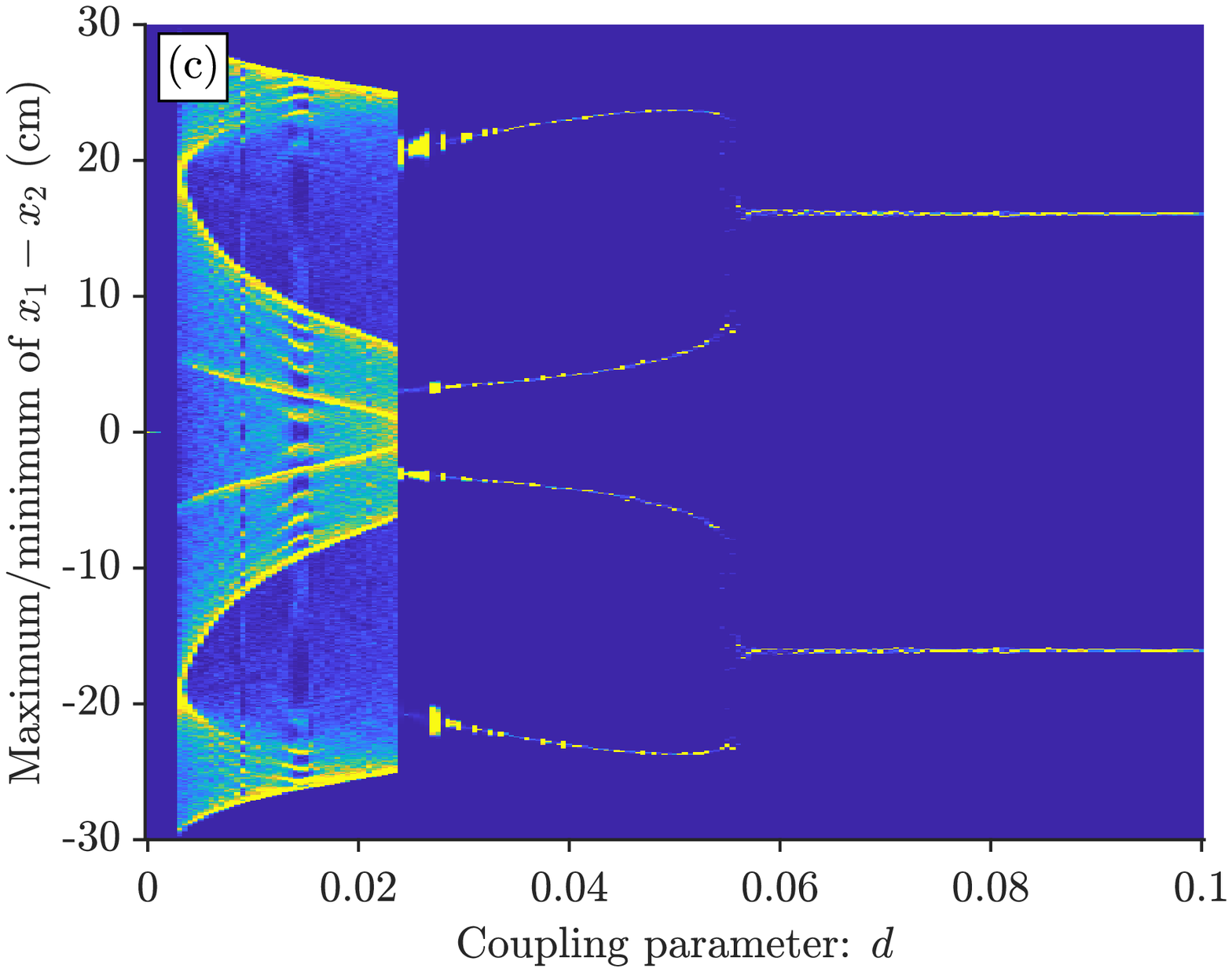}}
\subcaptionbox*{}{\includegraphics[scale=0.38]{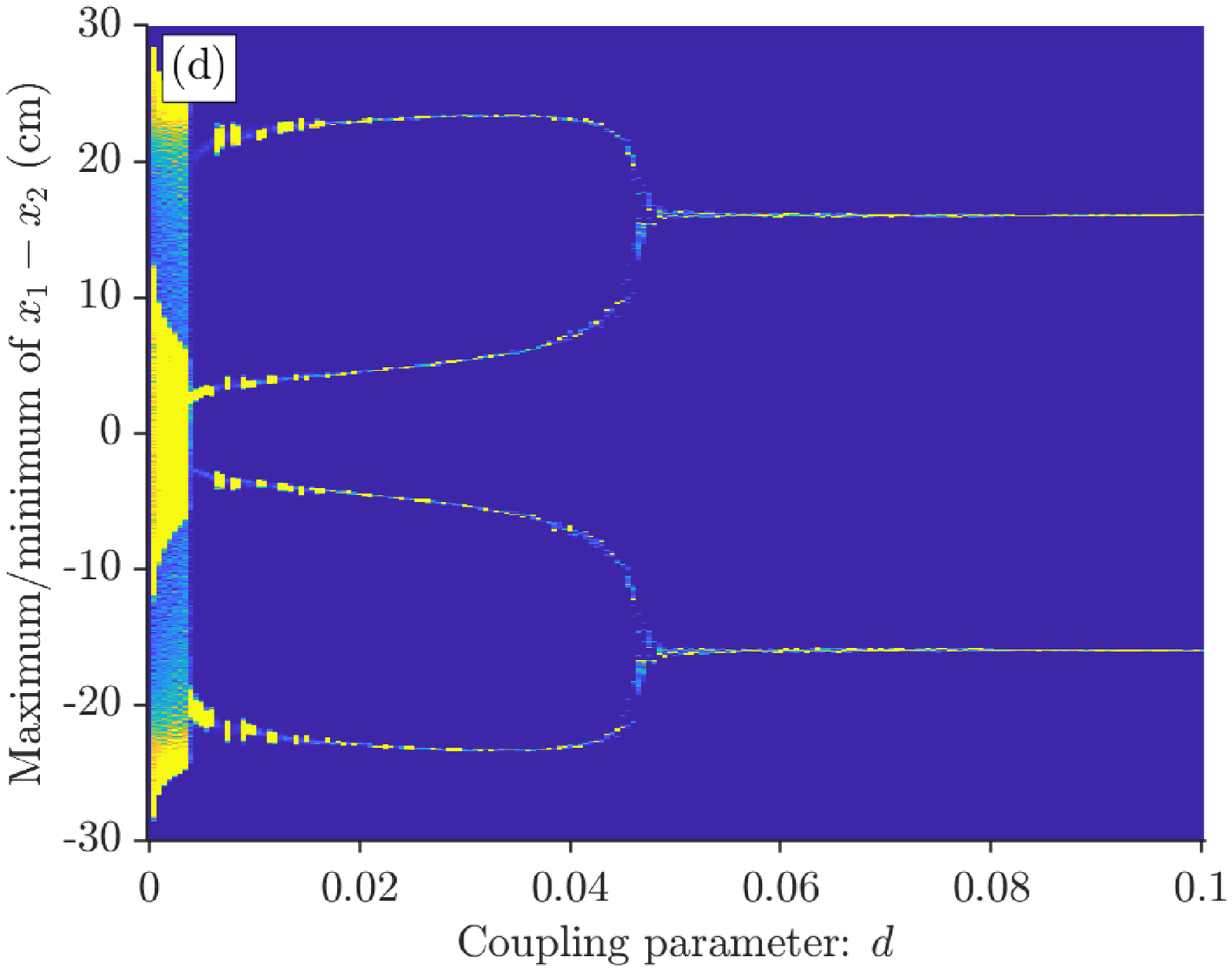}}
\caption{Heat plot bifurcation diagram with $d$ as bifurcation parameter for two fish model with $a = 15$,  $z_2 =0.15$,  $x_0=15$,  $b_0= 0$, $z_1 =50$ and $k= 0.4$;.  In (a) $c=0.92$, (b) $c=0.9255$,  (c) $c=0.9265$ and (d) $c=0.95$. The bifurcation parameter $d$ takes values between 0 and 0.1, with step size 0.0005. For each value of $d$ we simulate 10 times for 20,000 steps, each time with initial conditions  $(b_1(0), b_2(0), v_1(0), v_2(0),x_1(0), x_2(0))=(0,0,0,0,x_1^{start}, 0)$,  where $x_1^{start}$ is a random number between -20 and 20 cm.  The distance between the fish is plotted in 701 bins, meaning that each bin is of size 0.1 cm.  This allows us to create a two dimensional histogram with $701 \times 201$ bins. } \label{fig:twofish_bif_d}
\end{figure}

To get further understanding of how the model is affected both by the coupling parameter $d$, and the external input parameter $c$, we also looked at `heat map' bifurcation plots with $d$ as a bifurcation parameter, for different values of $c$ (figure \ref{fig:twofish_bif_d}).  Each subplot is produced by, for each value of $d$,  simulating the dynamical system 10 times, with random initial position for fish 1. For each iteration, find the maximum points of the difference in distance, $M_n$ and the minimum points of the difference in distance, $N_n$.  These are stored in a histogram with values ranging from -35 to 35 cm, from which we make a heat plot of this two dimensional histogram, where colour indicates how frequent a value is. 

In Figure \ref{fig:twofish_bif_d} (a), we see that when $c=0.92$, the fish move with constant speed and with no separation between them until $d \approx 0.058$, after which the system displays aperiodic leadership switching until $d \approx 0.07$, where there is no leader or follower, and instead the fish take turns in bursting and gliding.  Increasing $c$ to 0.9255 (Figure \ref{fig:twofish_bif_d} (b)),  the fish move with constant speed and with no separation in distance between them until $d \approx 0.0085$,  when they instead display aperiodic leadership switching.  When $d$ exceeds 0.017, the system shifts to periodic leadership switching, which is retained until $d \approx 0.0325$. For $0.0325<d<0.058$, there is always one leader and one follower, where the leading fish is dependent on the initial conditions.  After that, the there is no leader or follower, instead the fish has swimming dynamics as shown in Figure \ref{fig:twofish_timeseries} (i)-(j).  Increasing $c$ to 0.9265,  the system goes immediately from no separation in distance to periodic leadership switching when $d \approx 0.0035$.  After that,  there is one leader-one follower dynamics, where the  leading fish is dependent on the initial conditions, until $d=0.53$,  after which there is no leader or follower.  When $c=0.93$,  the one fish model has passed the bifurcation point. Thus, there is always burst and glide swimming, no matter the value of $d$.  For $d \in [0, 0.005]$ there is periodic leadership switching and after that the system exhibits one leader-one follower dynamics until $d=0.0445$, where there is no leader or follower. 

From all these bifurcation diagrams we see that there is a quite narrow range of values for the coupling parameter that give rise to interesting leader-follower dynamics -- when $c=0.9255$ all four behaviours are found within the interval $[0.008, 0.065]$ of the coupling parameter $d$ (Figure \ref{fig:twofish_bif_d} (b)).  The range of values for the external input parameter $c$ is even narrower: when $c$ is decreased by just 0.0055 from $c=0.9255$ to $c=0.92$,  the bifurcation diagram in Figure \ref{fig:twofish_bif_d} (a), shows that the the model does not display one leader-one follower dynamics for any values of $d$, and when $c$ is increased by just $0.0045$ from $c=0.9255$ to $c=0.93$, the bifurcation diagram in Figure \ref{fig:twofish_bif_d} (d) shows that the fish do not exhibit chaotic leadership switching.

\subsection{Quadratic drag}\label{quadraticdrag}

Many design choices were made when developing the model, but the main alternative assumption that requires further investigation is the effect of having velocity decrease in proportion to $v^2$. This is the natural choice from a hydrodynamical perspective, in particular for large fish. We now examine the model where both terms in the equation for $dv/dt$ are squared. This provides the same solution for the nullcline ($dv/dt=0$), and we have the following set of equations: 

\begin{align} \label{eq_onefish_v2_squared}
&\frac{db}{dt}=b-\frac{b^3}{3}+c - v \\
&\frac{dv}{dt}=a^2\left(\frac{\arctan(z_1 (b-b_0))}{\pi} +\frac{1}{2}\right)^2-(kv)^2.
\end{align}

\begin{figure}[H] \centering
\subcaptionbox*{}{\includegraphics[width=0.48\textwidth]{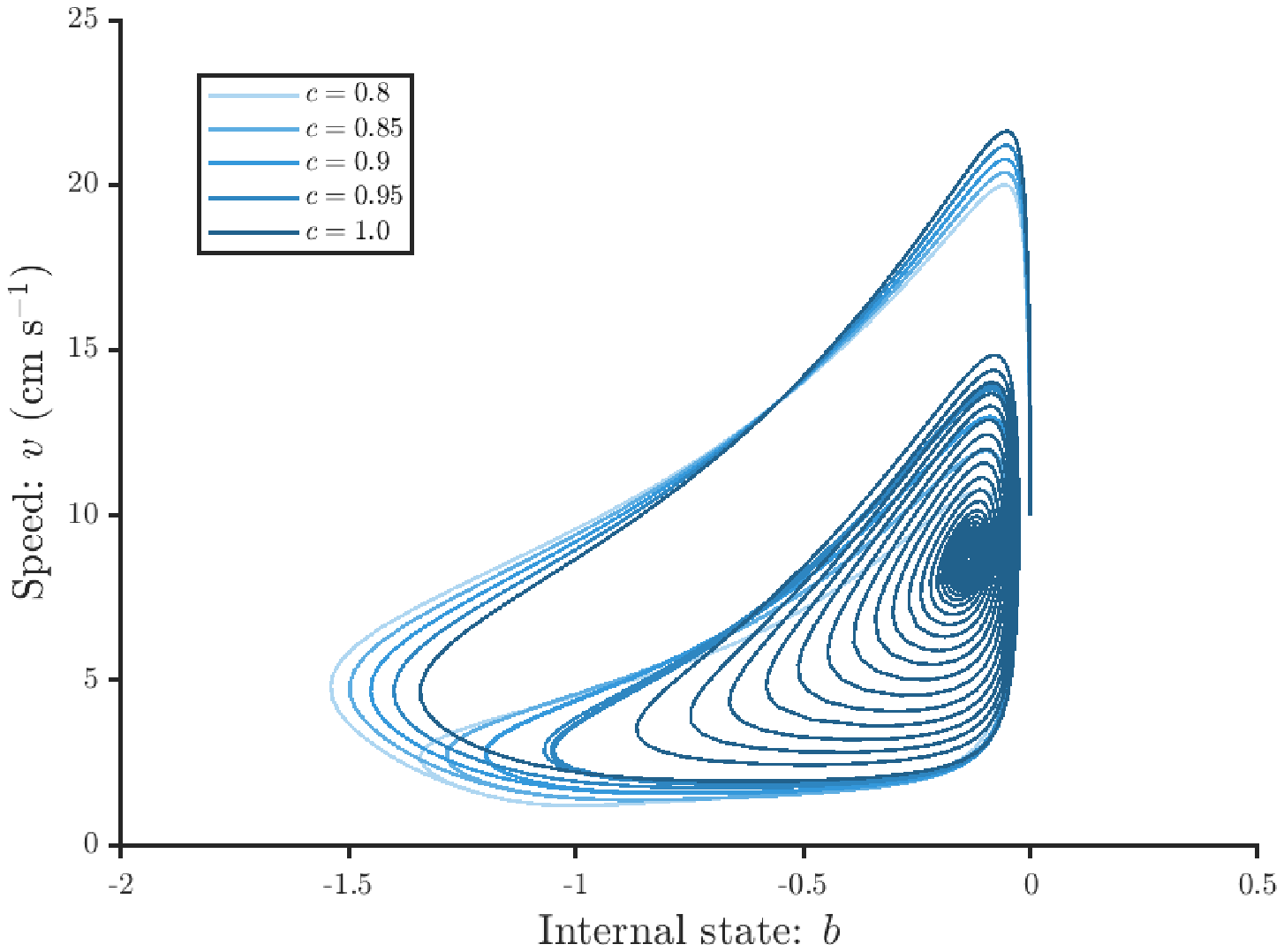}}
\subcaptionbox*{}{\includegraphics[width=0.48\textwidth]{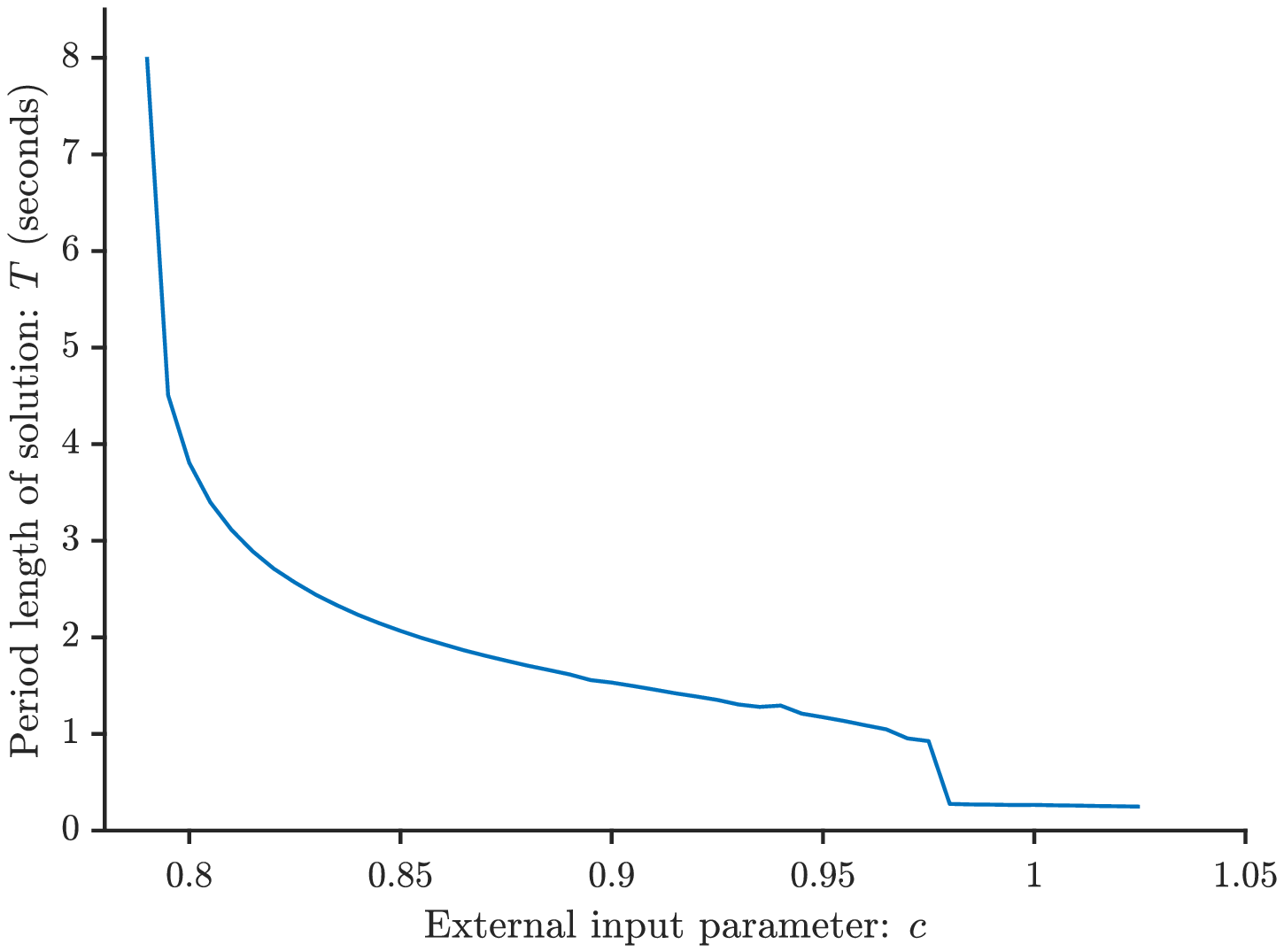}}
\caption{One fish model with $k=0.8$,  $a=15$ cm$\cdot$s$^{-2}$,  $z_1=50$, $b_0=0$ and quadratic drag where both terms in the equation for $dv/dt$ are squared.  Figure (a) shows the phase plane for four different values of $c$, as shown in the legend.  For $c\lesssim 0.79$ the dynamical system has a stable equilibrium,  resulting constant speed. For $c\gtrsim 0.79$ there is a stable limit cycle. As $c$ increases the amplitude of the speed peaks increases, until $c\gtrsim 0.98$ where the oscillation decays to a small circular limit cycle. Panel (b) shows the period length as a function of $c$. Increasing $c$ leads to decreasing period length and thus increasing frequency. } \label{fig:onefish_quadratic}
\end{figure}

In order to obtain a similar range of speeds, the speed decay was doubled to $k=0.8$. While the nullclines are now identical, the phase-space trajectories are still of a different shape to the linear case, as can be seen in Figure \ref{fig:onefish_quadratic}(a). There is a threshold for bursting for a single fish at $c \approx 0.79$, after which the bursting period decreases monotonically, until at $c\approx0.98$ where a canard explosion takes place \cite{krupa2001relaxation}.

\begin{figure}[H] \centering
\subcaptionbox*{}{\includegraphics[width=1\textwidth]{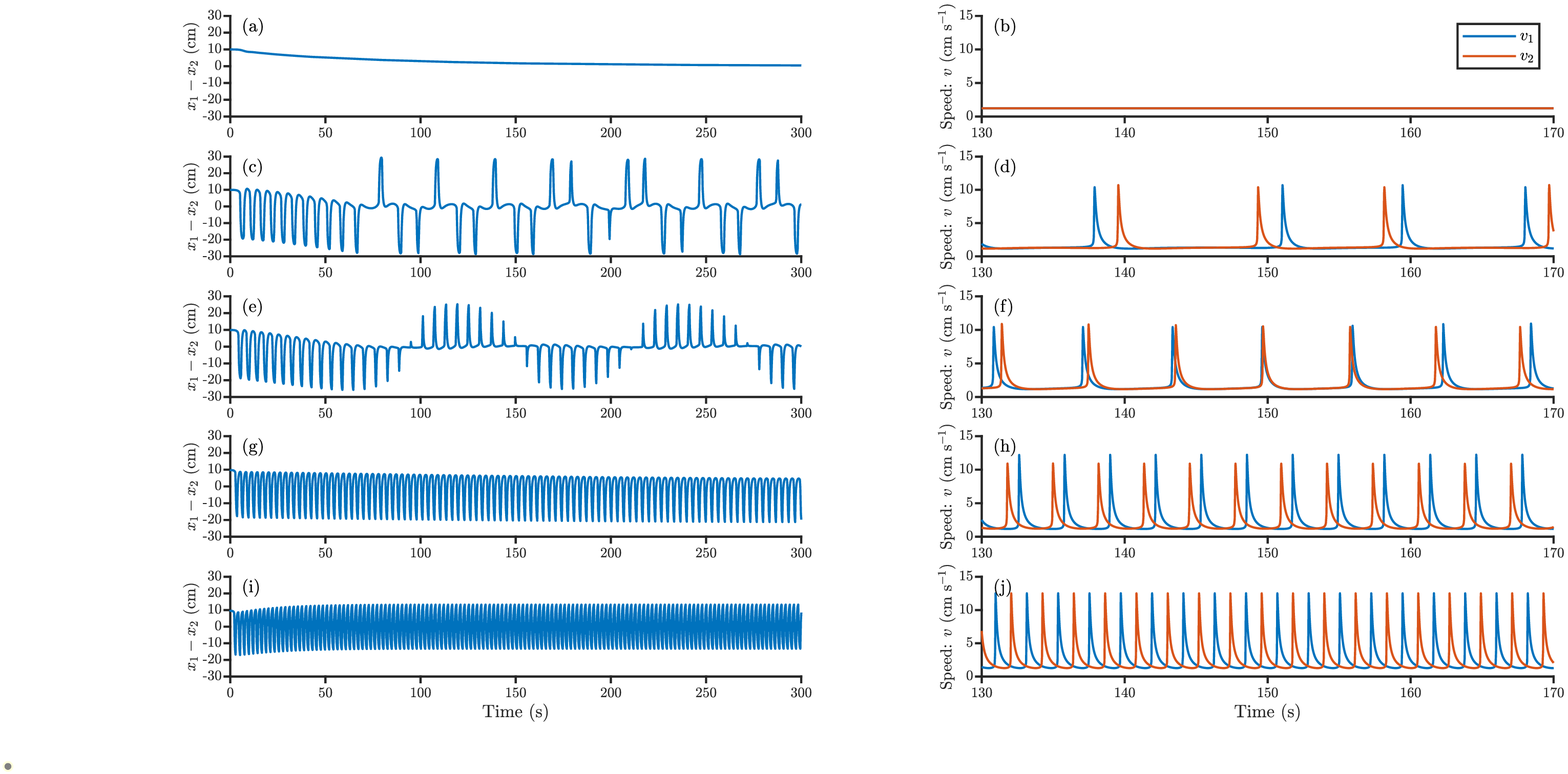}}
\caption{Two fish quadratic-drag model with $a = 15$,  $c=0.7885$,  $z_2 =0.15$,  $x_0=15$,  $b_0= 0$, $z_1 =50$ and $k= 0.8$.  In (a) and (b) where $d=0.0025$, after one speed burst each, the fish move with constant speed and stay exactly parallel; in (c) and (d) where $d=0.0125$, there is aperiodic leadership switching,  in (e) and (f) where $d=0.025$, there is periodic leadership switching, and in (g) and (h) where $d=0.1$, there is always one fish that is the leader,  and in (i) and (j) where $d=0.2$, there is no leader.} \label{fig:twofish_quadratic}
\end{figure}

For interacting pairs, the same sequence of behaviours as the linear drag case is observed as $c$ increases, as shown in Figure \ref{fig:twofish_quadratic}. However, the speed bursts are clearly sharper than for the linear case.

Another design choice we made, which is worth mentioning here, is the choice of arctan function for $g(b)$. This appears to be a necessary choice to get the dynamics we report here. When we tried different sigmoid functions for $g(b)$, e.g. a logistic function, the two fish model did not display the rich dynamics described in Section \ref{sec:twofish}. 

\section{Discussion}

The primary contribution of this paper is to show that burst and glide behaviour in fish can be reproduced by a burst and recover model of neurons. The fish velocity is viewed as being analogous to the recovery variable in the FitzHugh-Nagumo model for neural dynamics, and the internal state is analogous to the membrane potential variable.  The model reproduces, for different parameter values, many aspects of the intermittent locomotion found in many species of fish \citep{kalueff2013towards,wu2007kinematics, herbert2017predation, videler1982energetic,li2021burst}. In particular, for low coupling the fish move with a constant speed.  For slightly higher values the coupling parameter, $d$, the fish move by burst and glide swimming and switch leadership irregularly.  When increasing the strength of the coupling further,  the fish switch leaders at regular intervals. For even stronger coupling there is always one leader, with the leader depending on the initial distance between the fish, similar to that seen in pairs of eastern mosquitofish \cite{schaerf2021statistical}.   These results can be replicated with quadratic drag, but the burst periods are much steeper, which makes the model less biologically realistic for guppies, but might capture the interactions of larger fish species. 

In three of the four distinct forms of leader-follower dynamics, the follower fish bursts around a second after the leader (with the exact timing depending on parameter values), and has a higher maximum speed than the leader. However, the distance swum within one burst-and-glide period by the follower is always (slightly) shorter or the same distance as the distance swum by the leader within one burst-and-glide period. These qualitative model predictions would be relatively easy to test and may provide support for our model. For even higher values of the coupling parameter the fish alternate positions and have the same maximum speed and distance travelled within one burst and glide period. When one fish reaches its minimum speed, the other reaches its maximum speed.  

Another testable prediction arising for our model lies in the shape of the response function $f$. Previous experimental studies suggest that the response in speed is symmetric in the position of the other fish: the fish respond with a high speed both when the other fish is in front and behind \cite{kotrschal2020rapid}.  As seen in Figure \ref{fig:twofish_speedphase} (b), our model reproduces the same results but without that assumption.  This emphasizes the contributions of a mathematical model like ours: without a model,  the natural assumption from experimental data is that the bursting response is as strong when the other fish is behind than in front \cite{herbert2016understanding,gautrais2012deciphering,katz2011inferring}. However, we find that a highly asymmetric response function $f$, for which the fish responds less strongly to a conspecific behind it than in front, also gives rise to the sort of dynamics observed in experiments, when coupled with the delay induced by the `recovery' of the neuronal firing.    
This finding is reminiscent of the result of Perna et al., who showed that (in models) two very different response functions can give rise to the same behaviour \cite{perna2014duality}. In our case, the recovery process gives the impression that the fish in front is responding to the fish behind, while in fact it has simply completed a burst and glide cycle, and is ready to burst again.

Existing models of leadership switching in fish include a random component to allow the switching to happen. For example,  Markov chain models are often used to match experimental data of leadership switching \cite{harcourt2009social, nakayama2012initiative,  nakayama2012temperament}. Our model demonstrates that  leadership switching can occur in a purely deterministic interaction. The irregular leadership switching in our model apparently results from a form of deterministic chaos. The Lorenz map (Figure \ref{fig:twofish_lorenz} (a)) shows a clear deterministic form, while the autocorrelation indicates rapid decay in the correlation of movements. Evidence for chaos in animal behaviour has previously been presented in, for example, activity of single \textit{Leptothorax allardycei} ants \cite{cole1991animal} and colonies of foraging ants \cite{li2014chaos}. 

Chaotic dynamics is a crucial part of neural mechanisms \cite{korn2003there,  guevara1983chaos, skarda1987brains, wright1996dynamics,freeman1986eeg, rapp1985dynamics} and there are existing models connecting sensory systems to neuronal dynamics \cite{kuniyoshi2006early}. Our model appears to be the first in which chaos comes explicitly from a social interaction in a perception-action loop, rather than internally from the neurons. Chaos arises from the coupling between the intake of visual stimuli and the internal dynamics of the fish response.  There are increasing experimental studies linking neuronal activities in the brain to locomotion in fish \cite{naumann2016whole, dunn2016brain, chen2018brain}. Hopefully, our model can be used as a starting point for modelling this linkage.

While leadership switching is reasonably widespread in Nature, in our model it occurs when the coupling $c$ is close to the bifurcation point of the one fish model. Indeed, there is a very narrow range of parameters where the fish movements are chaotic. If leadership switching is ubiquitous in Nature, then one explanation for this might be an evolution of the neural parameter toward the Hopf bifurcation point at which burst and glide behaviour emerges. Similar mechanisms have been suggested in mathematical models describing foraging of ants and decision making slime mould \cite{nicolis2013foraging,  zabzina2014symmetry}. The argument in this case is that being near to a bifurcation point offers flexibility in foraging.  Similarly, it has been argued that biological chaos is generally favoured by selection \cite{ferriere1995chaos}. In our case, such an argument would suggest that evolutionary pressure takes fish responses towards the chaotic switching regime, where the fish regularly switch leadership. 

While burst and glide swimming is pivotal in single fish and pairs of fish, intermittent locomotion becomes less evident in schools of fish. Instead, leadership becomes less apparent and fish move more smoothly and are synchronized when schooling.  A future direction of our work would thus be to understand how this transition from turn-taking to coordinated movement occurs as the number of fish in a shoal increases.

\section*{Acknowledgements}

Thank you to Torbjörn Lund, Ernest (Liu Yu) and Maria-Rita D'Orsogna
 for a series of interesting discussions about modelling fish burst and glide and to the Mittag Leffler Institute for hosting those discussions. Thank you to James Herbert-Read for sharing the experimental data and explaining how the experiments were conducted. Thank you to Niclas Kolm for input on the biology and to Annika Boussard for showing Linnéa the guppies in the Kolm lab. 

\newpage
\bibliography{references_fish}

\end{document}